\documentclass[iop]{emulateapj}

\usepackage{amsmath}
\usepackage{subfigure}
\usepackage{url}
\usepackage{hyperref}
\usepackage{xspace}
\def\cloudy{\mbox{\rm{C\textsc{loudy}}}\xspace}
\def\sigame{\mbox{\rm{S\textsc{\'{i}game}}}\xspace}
\def\simba{\mbox{\rm{\textsc{simba}}}\xspace}
\def\mufasa{\mbox{\rm{\textsc{mufasa}}}\xspace}
\def\gizmo{\mbox{\rm{\textsc{gizmo}}}\xspace}
\def\caesar{\mbox{\rm{\textsc{caesar}}}\xspace}
\def\starburst{\mbox{\rm{\textsc{starburst99}}}\xspace}
\def\yt{\mbox{\rm{\textsc{yt}}}\xspace}
\def\msun{\mbox{\rm{M}$_{\odot}$}\xspace}
\def\lsun{\mbox{\rm{L}$_{\odot}$}\xspace}
\usepackage{color}

\def\gs{\mathrel{\raise0.35ex\hbox{$\scriptstyle >$}\kern-0.6em \lower0.40ex\hbox{{$\scriptstyle \sim$}}}}
\def\ls{\mathrel{\raise0.35ex\hbox{$\scriptstyle <$}\kern-0.6em \lower0.40ex\hbox{{$\scriptstyle \sim$}}}}
\newcommand{\Msolar}{\mbox{$\rm M_{\odot}\,$}}
\newcommand{\Lsolar}{\mbox{$\rm L_{\odot}\,$}}

\newcommand{\cii}{[C{\scriptsize II}]\xspace}
\shorttitle{Tracing gas in $z\sim 6$ galaxies}
\shortauthors{Vizgan et al.}

\begin{document}

\title{Tracing Molecular Gas Mass in $\MakeLowercase{z} \simeq 6$ Galaxies with \cii \\}

\author{D. Vizgan\altaffilmark{1,2,3,4}}

\author{T. R. Greve\altaffilmark{3,4,5}}

\author{K. P. Olsen\altaffilmark{6}}

\author{A. Zanella\altaffilmark{7}}

\author{D. Narayanan\altaffilmark{3,8}}

\author{R. Dav\`e\altaffilmark{9,10}}

\author{G. E. Magdis\altaffilmark{3,4,11}}

\author{G. Popping\altaffilmark{12}}

\author{F. Valentino\altaffilmark{3,11}}

\author{K. E. Heintz\altaffilmark{3,11,13}}

\altaffiltext{1}{Astronomy Department and Van Vleck Observatory, Wesleyan University, 96 Foss Hill Drive, Middletown, CT 06459, USA}
\altaffiltext{2}{Department of Astronomy, University of Illinois at Urbana-Champaign, 1002 West Green St., Urbana, IL 61801, USA}
\altaffiltext{3}{The Cosmic Dawn Center}
\altaffiltext{4}{National Space Institute, DTU Space,  Technical University of Denmark, 2800 Kgs.~Lyngby, Denmark}
\altaffiltext{5}{Dept.~of Physics and Astronomy, University College London, Gower Street, WC1E6BT London}
\altaffiltext{6}{Department of Astronomy and Steward Observatory, University of Arizona, Tucson, AZ 85721, USA}
\altaffiltext{7}{Istituto Nazionale di Astrofisica (INAF), Vicolo dell'Osservatorio 5, I-35122 Padova, Italy}
\altaffiltext{8}{Department of Astronomy, University of Florida, Gainesville, Florida 32611}
\altaffiltext{9}{Institute for Astronomy, University of Edinburgh, Edinburgh EH9 3HJ, U.K.}
\altaffiltext{10}{University of the Western Cape, Bellville, Cape Town 7535, South Africa}
\altaffiltext{11}{Niels Bohr Institute, University of Copenhagen, Lyngbyvej 2, DK-2100 Copenhagen, Denmark}
\altaffiltext{12}{European Southern Observatory, 85478 Garching bei M\"unchen, Germany}

\altaffiltext{13}{Centre for Astrophysics and Cosmology, Science Institute, University of Iceland, Dunhagi 5, 107 Reykjav\'ik, Iceland}

\begin{abstract}


We investigate the fine-structure \cii line at $158\,\mu$m as a molecular gas tracer by analyzing the relationship between molecular gas mass ($M_{\rm mol}$) and \cii line luminosity ($L_{\rm [CII]}$) in 11,125 $z\simeq 6$ star-forming, main sequence galaxies from the \simba simulations, with line emission modeled by \sigame{}. Though most ($\sim 50-100\,\%$) of the gas mass in our simulations is ionized, the bulk ($> 50\,\%$) of the \cii emission comes from the molecular phase. We find a sub-linear (slope $0.78\pm 0.01$) $\log L_{\rm [CII]}-\log M_{\rm mol}$ relation, in contrast with the linear relation derived from observational samples of more massive, metal-rich galaxies at $z\ls 6$. We derive a median \cii-to-$M_{\rm mol}$ conversion factor of $\alpha_{\rm [CII]} \simeq 18\,{\rm M_{\rm \odot}/L_{\rm \odot}}$. This is lower than the average value of $\simeq 30\,{\rm M_{\rm \odot}/L_{\rm \odot}}$ derived from observations, which we attribute to lower gas-phase metallicities in our simulations. Thus, a lower, luminosity-dependent, conversion factor must be applied when inferring molecular gas masses from \cii observations of low-mass galaxies. For our simulations, \cii is a better tracer of the molecular gas than CO $J=1-0$, especially at the lowest metallicities, where much of the gas is 'CO-dark'. We find that $L_{\rm [CII]}$ is more tightly correlated with $M_{\rm mol}$ than with star-formation rate (${\rm SFR}$), and both the $\log L_{\rm [CII]}-\log M_{\rm mol}$ and $\log L_{\rm [CII]}-\log {\rm SFR}$ relations arise from the Kennicutt-Schmidt relation. Our findings suggest that $L_{\rm [CII]}$ is a promising tracer of the molecular gas at the earliest cosmic epochs.

\end{abstract}


\section{Introduction}

 Molecular gas and star formation are observed to be tightly
 correlated \citep[e.g.][]{Kennicutt1998, kennicutt2012, bolatto2013} in our
 Galaxy as well as in local and high-redshift galaxies, with the possible
 exception of star-formation in extremely metal-poor environments where stars may
 form directly from atomic gas inflows \citep{Krumholz2012,Michalowski2015}. 
 Unfortunately, the dominant component of molecular gas, H$_2$, is invisible under the typical conditions
 prevailing in the molecular interstellar medium (ISM).  Due to its lack of a
 permanent dipole moment, H$_2$ only emits from warm ($> 100\,{\rm K}$) gas
 \citep{draine2011}, and its two lowest rotational transitions have upper
 energy levels of $510\,{\rm K}$ and $1015\,{\rm K}$.  The low ($< 10-50\,{\rm K}$)
 temperatures typical of the giant molecular clouds \citep{cox2005}, which make
 up the bulk of the star formation fuel, are unable to excite H$_2$ and are
 therefore '$\rm{H_2}$-dark' to observers. Our inability to directly trace the
 distribution and amount of molecular gas in galaxies is a long-standing challenge for
 extragalactic studies and has led to the development of a number of indirect
 H$_2$-tracing methods \citep[]{carilli2013, li2018}, which we briefly summarize below.

The $J=1-0$ rotational transition line of carbon monoxide (CO) is
 the most rigorously calibrated and most regularly used tracer of molecular gas
 mass \citep[e.g.,][]{Dickman1986,Solomon1991,Downes1998,narayanan2012, bolatto2013}.  The utility of the CO line comes
 from the fact that it is relatively bright and easily
 excited in molecular ISM conditions. 
 
 To convert the observed CO(1-0) line luminosity into a molecular gas mass estimate, a CO-to-$\rm{H_2}$
 conversion factor, $\alpha_{\rm CO}$, is needed. The conversion factor is
 calibrated for giant molecular clouds (GMCs) in our Galaxy and nearby galaxies \citep[e.g.,][]{bolatto2013}. A
 galaxy-averaged value for $\alpha_{\rm CO}$ has been determined for local
 IR-luminous galaxies and found to be $\sim 4\times$ lower than the Galactic
 value \citep{Downes1998}, albeit with significant scatter \citep{Papadopoulos2012}.  For high-$z$
 galaxies, we do not have strong constraints on $\alpha_{\rm CO}$ \citep[][]{tacconi2013, tacconi2018, valentino2020}; both
 observations and simulations suggest that $\alpha_{\rm CO}$ depends on
 the gas temperature,
 velocity dispersion, and metallicity \citep{Tacconi2008,Ivison2011}. In regions of high gas temperature and
 velocity dispersion, the optically thick CO(1-0) emission line emits more
 luminosity per unit H$_{\rm 2}$ column density, which leads to lower conversion
 factors as seen in the local IR-luminous galaxies.  In low-metallicity galaxies, $\alpha_{\rm
 CO}$ scales inversely with metallicity since in such environments, CO is
 photo-dissociated in a larger fraction of the molecular gas, which is thus
 referred to as 'CO-dark' gas.
 \citep[e.g.,][]{Wilson1995,Arimoto1996,Wolfire2010,narayanan2012,Genzel2012}.
 Intense cosmic-ray
 ionization rates that come along with extreme star-formation activity can
 destroy CO deep within molecular clouds, thereby further degrading its H$_2$-tracing
 capabilities \citep{Bisbas2015,Bisbas2017}. 
 Furthermore, at high redshifts, the cosmic
 microwave background (CMB) becomes brighter than the low-$J$ CO emission from
 cold molecular gas; for example, \cite{dacunha2013} showed that at $z \gs 3$, the CMB could significantly affect both CO(1-0) and CO(2-1) line fluxes,
 and neglecting these effects can lead to non-negligible underestimations of the
 derived luminosities and, therefore, the gas masses \cite[see also][]{Tunnard2016,Zhang2016}.
 
The fine-structure transitions of neutral carbon -- \textbf{[C{\sc i}]}($^3P_1$$-$$^3P_0$)
 and \textbf{[C{\sc i}]}($^3P_2$$-$$^3P_1$) have also been proposed as tracers of the molecular gas in galaxies \citep{Gerin2000,papadopoulos2004, narayanan2017}. Carbon is abundant ($\sim 10^{-5}$
 relative to H$_2$), yet the \textbf{[C{\sc i}]} lines have low to moderate optical depths
 ($\tau_{\rm \textbf{[C{\sc i}]}} \sim 0.1-1$). Both lines are easily excited in typical
 molecular gas conditions, which, together with its simple 3-level partition function, makes it a
 promising tracer of H$_2$.  \textbf{[C{\sc i}]} surveys of molecular clouds in our
 Galaxy \citep{Ikeda2002} showed a remarkable spatial and kinematical
 similarity in the \textbf{[C{\sc i}]} and CO emission, suggesting that \textbf{[C{\sc i}]} does indeed
 trace the bulk H$_2$ reservoir. Simulations of molecular
 clouds have shown that cloud inhomogeneity (resulting in deeper UV penetration
 and thus an increase in the \textbf{[C{\sc i}]} layer; \citealt{Offner2014}) and turbulence
 (smoothing any initial \textbf{[C{\sc i}]}/CO abundance gradient; \citealt{Glover2015}) are
 responsible for the observed \textbf{[C{\sc i}]}-CO concomitant. \citet{Offner2014}
 further find that over a wide range of column densities \textbf{[C{\sc i}]} is a
 comparable or superior tracer of the molecular gas, especially when an intense
 UV field is impinging on it. Another work on absorption lines in high-redshift galaxies suggests a universal metallicity-dependent \textbf{[C{\sc i}]}-to-H$_2$ conversion factor \citep[][]{heintz2020}. Recent work \citep[][]{valentino2018, valentino2020b} has also investigated the utility of \textbf{[C{\sc i}]} as a gas mass tracer in the high-$z$ universe.
 
 The far-IR/sub-millimeter dust continuum emission has been used as an indirect
 tracer of the total gas content (molecular and atomic) of the Galactic ISM
 \citep{Hildebrand1983}, of nearby galaxies
 \citep{Guelin1993,Guelin1995,Eales2012}, and of high-$z$ galaxies
 \citep{Magdis2012,Scoville2014,Scoville2016,liang2018,privon2018, Kaasinen2019}.  One approach has been
 to derive the dust mass from careful modelling of the far-IR/sub-millimeter
 spectral energy distribution (SED), and apply a dust-to-gas mass ratio. The
 problem with this method is not only the uncertainties associated with the SED
 fitting but a number of other factors, such as the dust temperature, the grain
 emissivity, and the dust-to-gas mass conversion factor. The latter depends on metallicity \citep[][]{remyruyer2014, li2019},
 which implies a redshift dependence via the mass-metallicity relation.
 Furthermore, this method requires flux measurements over a range of
 far-IR/(sub-)mm bands. To circumvent these problems, a direct calibration
 between the gas mass and the dust continuum emission at a single, broadband
 (sub-)mm wavelength has been sought \citep[e.g.,][]{Scoville2014}. Such
 calibrations have been established for optically thin continuum emission along
 the Rayleigh-Jeans tail, e.g., at $850\,{\rm \mu m}$, and successfully applied
 to both local and high-$z$ galaxies \citep{Scoville2014,Scoville2016}. This
 method has the advantage over CO and \textbf{[C{\sc i}]} in that it is easier to detect the
 continuum emission in a single broadband than it is to detect a single line.
 At redshifts $\gs 4-5$, however, this method too runs into the limitations set
 by the increasing CMB temperature, which will not only couple thermally to the
 cold dust but also 'drown out' a significant fraction of the dust continuum
 emission \citep[e.g.][]{dacunha2013,Zhang2016}.

 In recent years, there has been a growing effort, from both the observation and simulation sides, to explore the $158\,{\rm \mu m}$ ($1900.5\,{\rm
 GHz}$) fine-structure transition of singly-ionized carbon ([C{\sc ii}]) as a molecular gas
 mass tracer \citep[][]{Accurso2017,hughes2017a,zanella2018,dessauges2020,madden2020}
 The [C{\sc ii}]158${\rm \mu m}$ line is one of the strongest cooling lines of
 the ISM, and can carry up to a few per-cent of the total far-IR energy emitted
 from galaxies \citep[e.g.,][]{Malhotra1997}. Carbon is the fourth most abundant element and has an ionization potential of $11.3\,{\rm
 eV}$, lower than that of neutral hydrogen. \cii, therefore, permeates much of
 the ISM. It is found in photodissociation regions (PDRs), diffuse ionised and
 atomic regions, and even molecular gas \citep[e.g.,][]{Kaufman1999}.  
 Because the [C{\sc ii}] line arises in most phases of the ISM, it is
 important to disentangle the different contributions to
 interpret galaxy-wide, integrated [C{\sc ii}] observations correctly. 
 How much of the [C{\sc ii}] emission comes from the molecular phase, and to what
 extent the line can be used as a molecular gas tracer is of particular interest.
 Observations of \cii in the Milky Way, for instance, show that $\sim$75\% of the \cii
 emission comes from dense PDRs and CO-dark H$_2$ gas
 \citep{pineda2014}. 

Because of its association with PDRs, \cii has been viewed as a tracer of the star-formation rate (SFR) \citep[e.g.][]{delooze2014,magdis2014,herreracamus2015,schaerer2020}. At low redshift, metal-poor dwarf galaxies and star-forming galaxies display slightly offset, but tight log-linear relations between the \cii luminosity and the sum of total obscured and unobscured SFR \citep{delooze2014}. However, compact starbursts with high SFR surface densities and intense UV radiation fields tend to exhibit a '\cii/FIR deficit' by as much as an order of magnitude compared to the local \cii-SFR relations \citep[e.g.][]{magdis2014, diazsantos2013, narayanan2017}. The \cii/FIR deficit is typically observed for local ultra-luminous IR galaxies (ULIRGS; $L_{\rm IR} \geq 10^{12}\,{\rm \Lsolar}$) and high-$z$ starbursts/mergers, while normal star-forming galaxies at high-redshifts are in line with the local \cii-SFR relationship \citep[e.g.,][]{Carniani2018}. \citet{schaerer2020} also found that there is little evolution in the \cii-SFR relationship for main-sequence galaxies both locally and in the high-redshift (i.e., $4 < z < 8$) universe, albeit the scatter is larger at higher redshift.

 By extension, this explains why \cii
 traces molecular gas since the Kennicutt-Schmidt law shows a strong linear
 relationship between SFR and molecular gas mass. However, this relationship is
 not unique; from observations, normal galaxies seem to differ from
 starbursts/mergers in that the latter seem to deplete their gas reservoirs
 around 10 times faster. 
 As pointed out by \citet{zanella2018}, if one assumes \cii traces molecular gas mass, the \cii/IR
 relationship becomes $M_{\rm mol}$/SFR, which is the gas depletion timescale of a
 galaxy. In this case, the observed 'FIR deficit' of starbursts reflects their $10\times$
 shorter gas depletion time-scales compared to normal star-forming galaxies. 
 Theoretically, a strong connection between \cii luminosity and molecular gas
 mass is also expected from the fact that the low critical density of
 carbon ($11.3\,$eV) allows for the excitation of \cii in dense,molecular regions
 by collisions with H$_2$ molecules. Indeed, simulations of high-redshift normal
 galaxies \citep[e.g.][]{vallini2015, popping2019, Leung2020} found that most \cii emission comes from dense PDRs associated with regions of molecular gas; Though \citet{olsen2017} found that most ($\sim 64\%$) of \cii emission arose from ionized gas clouds in their simulations, they amended this result in an erratum, after accounting for the CMB in \cloudy, and found that about 50\% of \cii emission arose from molecular gas clouds \citep[see][]{olsen2018}.
 
 The use of \cii as a molecular gas mass tracer has been implemented on a sample of 10
 main-sequence galaxies at $z\sim2$ by \cite{zanella2018} (henceforth; Z18). The
 authors combine Band 9 ALMA (Atacama Large Millimeter/submillimeter Array)
 observations of \cii emission with existing multiwavelength observations to
 derive an average \cii mass-to-luminosity conversion factor $\alpha = 31
 M_{\sun}/L_{\sun}$ with an uncertainty of 0.3 dex. The molecular gas mass was derived from the integrated
 Schmidt-Kennicutt relation using the SFR measured from pixel-by-pixel SED fitting, while an independent and consistent gas mass estimate was derived from the dust continuum.
 The resulting conversion factor was found to be mostly invariant with galaxies’
 redshift, depletion time, and gas-phase metallicity. At lower redshifts \cii has
 also been demonstrated to trace molecular gas mass \citep{madden2020}. Notably, \cii has also been shown to trace atomic hydrogen in galaxies until the epoch of reionization \citep[][]{heintz2021}.
 
 The aim of this paper is to investigate whether \cii is still a reliable tracer of the molecular gas
 mass for lower-mass and lower-metallicity galaxies at earlier epochs (i.e. z $\approx$ 6). Repeating the experiment on a much larger sample of simulated high-redshift
 galaxies, we aim to derive a conversion factor for the molecular gas mass of
 high-redshift galaxies using \cii line luminosities, as has been done in the
 nearby universe ($z \sim 0.03-0.2$) for CO \citep{hughes2017a}, and as done for
 galaxies in the local Herschel Dwarf Galaxy Survey for \cii
 \citep[see][]{madden2020}.

 In this paper we present \sigame (Simulator of Galaxy Millimeter/Submillieter Emission) simulations of \cii line emission from 11,125 galaxies at $z\simeq 6$ in order to examine whether the line can be used to trace the gas content in normal star-forming galaxies at cosmic dawn. 
 Section 2 of the paper describes the observation
 samples that we compare our simulations with in this work; Section 3 describes the simulation sample and the post-processing done by \sigame. 
 Section 4 compares the integrated physical properties of the simulation sample in the context of observations. 
 Section 5 examines \cii as a tracer of molecular gas mass and as a tracer of star formation rate, derives a \cii-to-H$_{\rm 2}$ conversion factor ($\alpha_{\rm \cii}$), and investigates the origin of the $M_{\rm mol} - L_{\rm \cii}$ relation.
 Section 6 summarizes the main conclusions from our work.
 
\section{Observed comparison samples }\label{section:comparison-samples}
We employ two observed galaxy samples for comparison with our simulations. The first is from Z18, who presented six \cii-detections out of a sample of ten $z \sim 2$ main sequence galaxies observed with ALMA (Atacama Large Millimetre/submillimetre Array), and combined these with other \cii detections from the literature. The latter spanned the redshift range $0 \le z \le 6$ and included diverse galaxy populations, from local dwarfs and normal star-forming galaxies to luminous starbursts. Independent estimates of their molecular gas masses for all of these galaxies were given in Z18, mostly from existing observations of CO but also via more indirect methods such as from parametrizations of the gas depletion time ($\tau_{\rm depl} = M_{\rm mol}/{\rm SFR}$) as a function of the specific star-formation rate (${\rm sSFR} = {\rm SFR}/M_{\rm \star}$). In this paper, we split the Z18 sample into $z \geq 4$ and $z < 4$ galaxies. 

We additionally compare our simulations to results obtained from the ALPINE-ALMA \cii Survey \citep[henceforth referred to as ALPINE; see][]{lefevre2020, dessauges2020}. ALPINE
has observed 118 galaxies in the redshift range $4  <  z  <  6$ for \cii and dust FIR emission. In 64\% of galaxies observed, \cii was detected. These 75 galaxies, with which we compare our simulations, consist mainly of main sequence galaxies, with only a handful of sources lying either above or below the main sequence at these redshifts. The molecular gas masses of the ALPINE sample have been estimated from their restframe $850\,{\rm \mu m}$ continuum luminosity extrapolated from restframe 158 $\mu$m continuum; see \citet{dessauges2020} for details. The SFRs of the ALPINE sample are derived, via Eq.~4 in \citet{bell2003}, from their IR luminosities, which were inferred from the IRX-$\beta$ relation \citep[see][]{fudamoto2020}. These $L_{\rm IR}$-derived SFRs were added to the UV-derived SFRs \citep[]{faisst2020} in order to get the total SFRs.

We use the Z18 and ALPINE samples in combination with our simulations (see Section \ref{section:simulations}) to examine \cii as a tracer of molecular gas.

\section{Simulations}\label{section:simulations}

\subsection{Hydrodynamic simulations}\label{subsection:simulations}
This work builds on the analysis of snapshots taken from the \simba suite of cosmological galaxy formation simulations, which themselves were evolved using the meshless finite mass hydrodynamics technique of \gizmo \citep{hopkins2015,hopkins2017,dave2019}. The \simba simulation set consists of three cubical volumes of 25, 50 and $100\,{\rm cMpc}\,h^{-1}$ on a side, all of which are used in this work to search for galaxies at $z\sim6$. For each volume, a total of 1024$^3$ gas elements and 1024$^3$ dark matter particles are evolved from $z=249$. Compared to its predecessor, \mufasa, new features in \simba include the growth and feedback of supermassive black holes as well as a subgrid model to form and destroy dust during the simulation run; for further details we refer to \cite{dave2019} and \cite{li2018} for these two processes, respectively. The galaxy properties in \simba have been compared to various observations across cosmic time \citep[][]{thomas2019, Appleby2020}, including the epoch of reionization \citep[][]{Wu2020, Leung2020}, and are in reasonable agreement.

Since version 2 of \sigame used here relies heavily on the molecular gas mass fraction in the simulated galaxies, we briefly describe how this fraction is derived in \simba, referring to \cite{dave2020} for a more detailed description. 
The molecular gas mass content of each fluid element in \simba is calculated at each time step following the subgrid prescription of \cite{krumholz2011}.
This prescription relies on local metallicity and gas column density, modified to account for variations in resolution \citep{dave2016}. 
The H$_{\rm 2}$ mass function (H2MF) was found be in good agreement with observations at redshifts $z\sim0-2$, albeit overpredicting the H2MF at the high mass end \citep{dave2020}. 

Importantly, when comparing the \cii luminosity derived with \sigame to the molecular gas content of the simulated galaxies, we will not be using the molecular gas mass fractions from \simba, but rather the re-calculated molecular gas masses from \cloudy, as described in the following section. This is because the molecular gas masses as calculated in \simba are rough estimates based on local gas densities ($\sim$ 1 $\times$ 10$^2$ cm$^3$) and metallicities within the simulations, whereas the post-processed molecular gas masses from \cloudy account for the much higher densities ($\sim$ 10$^5$ -- 10$^6$ cm$^3$) of the molecular clouds in the simulations. As a consistency check, we compared the \sigame vs \simba molecular gas masses and found, as expected, a strong correlation between the two quantities.

From the \simba simulations, 11,137 galaxies were selected via the yt \citep[][]{turk2011}-based package \caesar; \yt is a Python package used to analyze and visualize volumetric data, and \caesar is a six-dimensional friends-of-friends algorithm which is applied to the simulated gas and stellar particles to identify and select \simba galaxies. Our final sample consists of 11,125 galaxies between a \cii luminosity range of 10$^{3.82}$ to 10$^{8.91}$ \lsun{} and a molecular gas mass range of 10$^{6.25}$ to 10$^{10.33}$ \msun{} The galaxy properties derived with \caesar include star formation rate, computed by dividing the stellar mass formed over a 100 Myr timescale \citep[][]{Leung2020}, star formation rate surface density, stellar mass (M$_{\rm \star}$) and total gas mass. 
In addition, we derive a SFR-weighted gas-phase metallicity, $Z$. All of these properties will be used for the analysis in Section \ref{section:sample-properties}. 

\subsection{\sigame post processing}
The sample of \simba galaxies are post-processed with version 2 of the
\sigame module \citep{olsen2017}.\footnote{\url{https://kpolsen.github.io/SIGAME/index.html}} At
its core, this version of \sigame uses the spectral synthesis code \cloudy \citep[v17.01;][]{Ferland2013, Ferland2017} to model the line emission from the multi-phased ISM within
each simulated galaxy. Using physically motivated prescriptions, \sigame calculates, throughout the simulated galaxy, the local interstellar radiation field (ISRF) spectrum, the cosmic ray
(CR) ionization rate, and the gas density distribution of the ionized, atomic and molecular ISM phases. 
The local equilibrium gas temperatures are calculated for each ISM phase by balancing the gas heating and cooling.

\sigame calculates the local ISRF that impinges on a gas particle in the simulations as the sum of the radiation field from  stars within one smoothing length of that gas particle. The radiation from a stellar particle is the integrated spectrum from \starburst modeling of the stellar population contained in the stellar particle, based on its mass, age and metallicity. 
The local CR ionization rate is assumed to scale linearly with the local ISRF far-UV 
strength. As for the metal abundances, gas-phase metallicities are carried over
from the \simba simulations. Finally, the gas density requires a sub-gridding
of the fluid elements which are themselves too large ($\geq 10^5$\,\msun) to follow
the intricate processes taking place inside giant molecular clouds (GMCs)
of mass $10^4$--$10^6$\,\msun. This sub-gridding is carried out via subdivision
of the gas into three ISM phases. First, the gas mass of each fluid element is
divided into a dense and a diffuse part based on the H$_2$ gas mass fraction of
the simulation itself. The diffuse gas is further subdivided into neutral and
ionized diffuse gas, leading to three ISM phases; dense gas which is later on
organized in giant molecular clouds (GMCs), diffuse neutral gas (DNG) and
diffuse ionized gas (DIG). 
Finally, the level populations and line emission are derived with \cloudy, which also provides mass fractions of atomic, ionized and molecular hydrogen. 
It is from these mass fractions that we derive the final molecular gas mass of each galaxy. 

In general the molecular gas masses re-calculated in this manner do not exceed those of the original \simba simulation, as that sets the maximum mass of all GMCs, of which \cloudy will deem a certain fraction (close to $100\%$) to be molecular. 
For a full description of the main methods and
assumptions of \sigame, we refer to \cite{olsen2017} and references therein. For this paper, we have used a slightly updated version of \sigame and we refer to  \citet{Leung2020} for a description of these updates and the creation of the data set analyzed here.

\section{Sample properties}\label{section:sample-properties}
\subsection{Integrated properties}\label{subsection:integrated-properties}
\begin{figure}[h!]
    \centering
    \includegraphics[width=0.96\columnwidth]{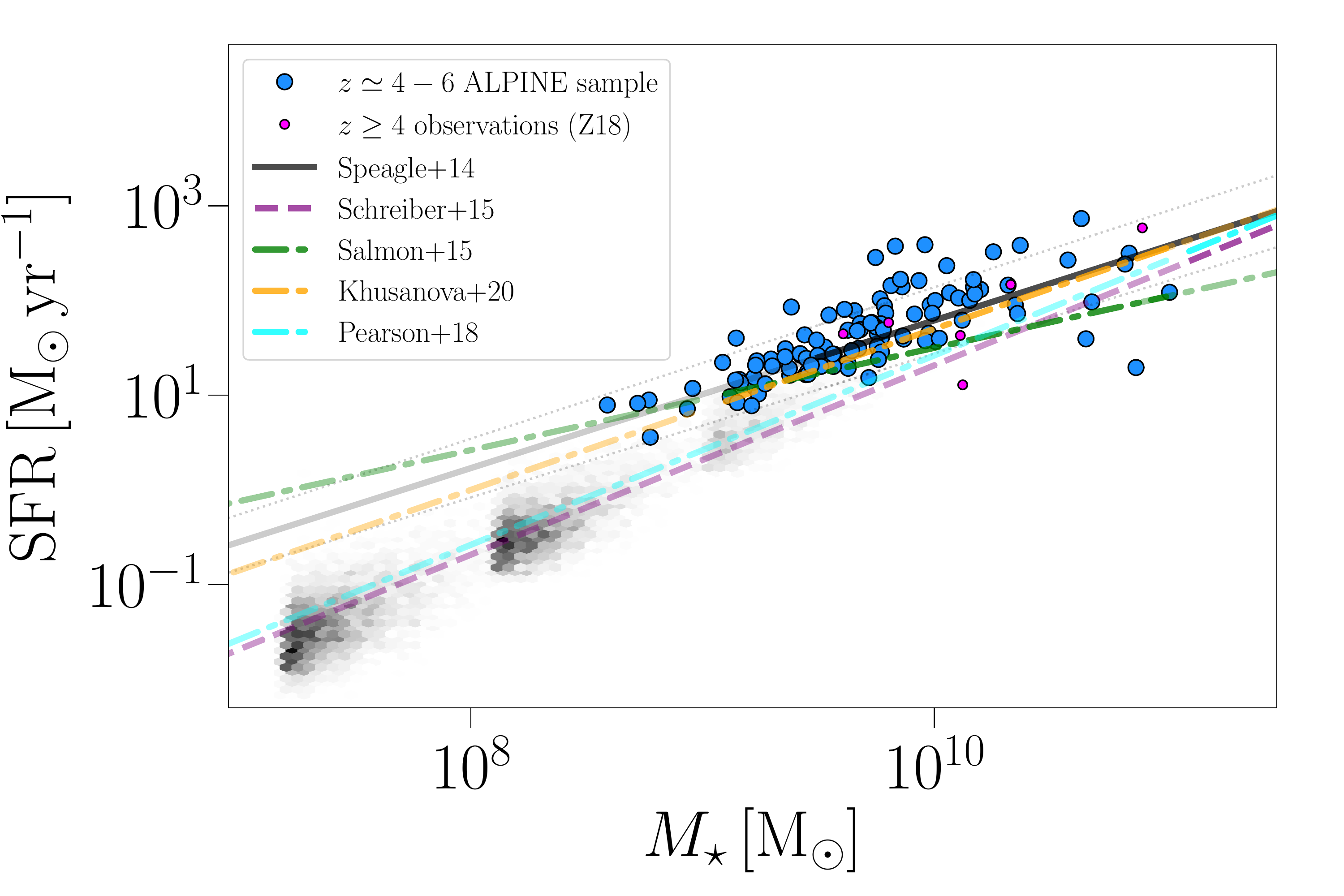}
    \includegraphics[width=0.96\columnwidth]{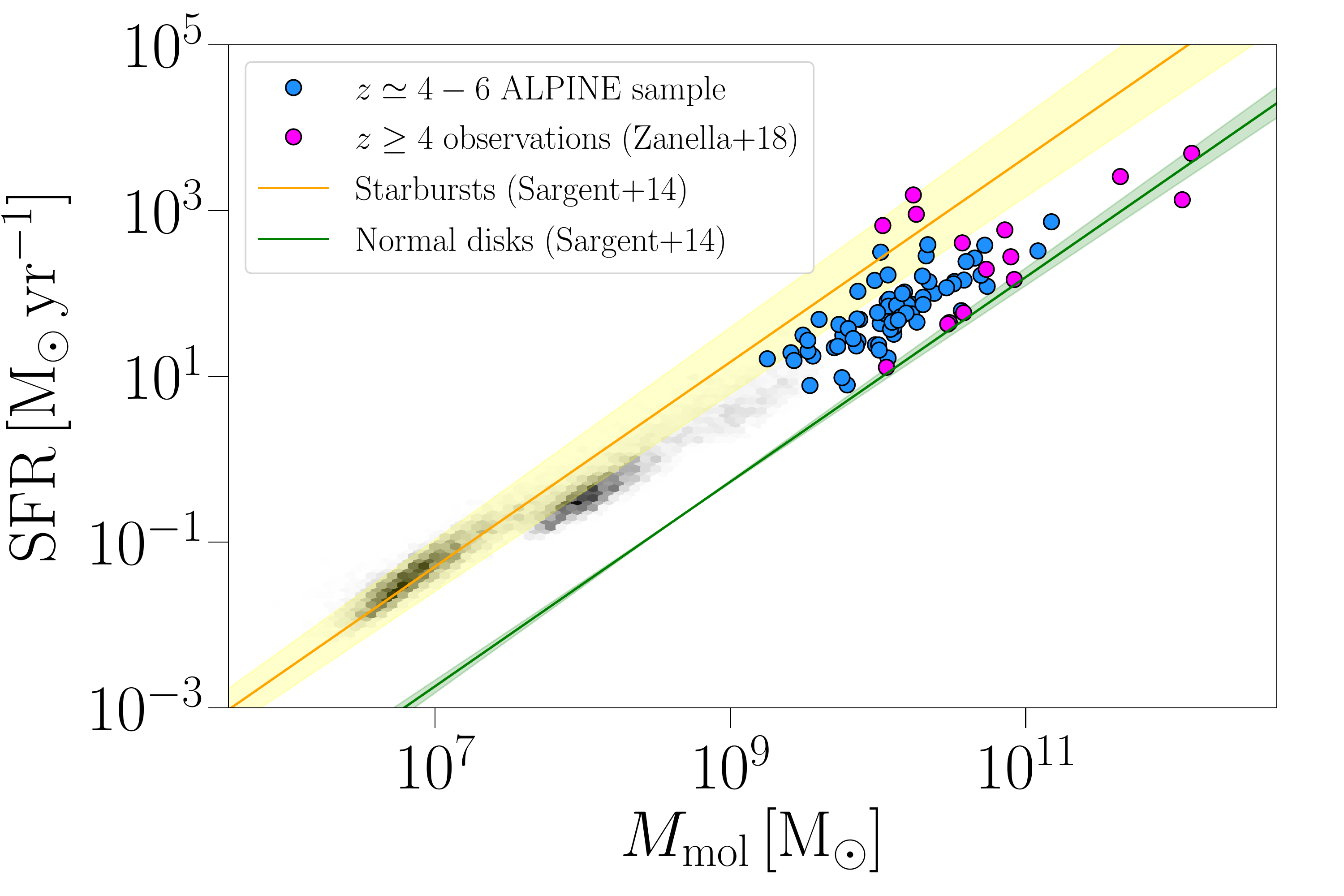}
    \includegraphics[width=0.96\columnwidth]{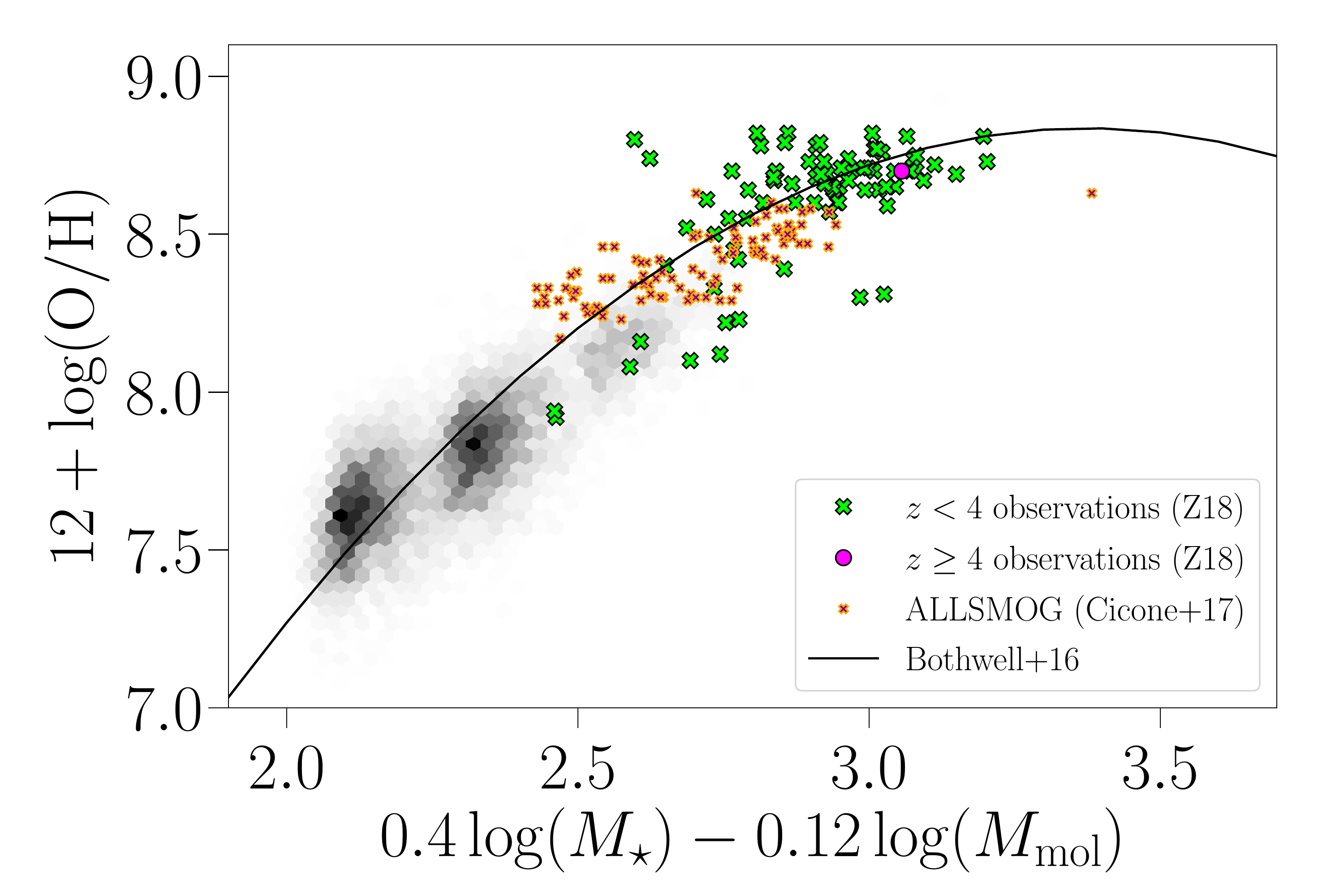}
    \caption{{\bf Top:} The ${\rm SFR}$ vs $M_{\rm \star}$ plane with the distribution of our
    simulated galaxies (grey hexbins) shown, along with the ALPINE sample (blue filled circles) and the $z > 4$ galaxies from Z18 (magenta filled circles). The lines show best-fit power-laws to observations of the galaxy main sequence at $z\sim 5-6$ from \cite{speagle2014} (solid black lines, with 1-$\sigma$ scatter indicated by dotted lines), \cite{Schreiber2015} (dashed purple), and \cite{Salmon2015} (dot-dashed green). The highlighted segments of the lines indicate the stellar mass ranges of the galaxy samples used to infer the main sequence. {\bf Middle:} The integrated Kennicutt-Schmidt relation (${\rm SFR}$ vs $M_{\rm mol}$) for our simulated and the ALPINE and $z > 4$ Z18 samples. Where they overlap, the simulations and observations follow the same relation although the latter exhibit a larger scatter. {\bf Bottom:} The molecular fundamental mass-metallicity relation, i.e., metallicity ($12 + \log ({\rm O/H})$) vs the optimum linear 
 combination of $\log M_{\rm \star}$ and $\log M_{\rm H_2}$ as determined from 
 a PCA analysis of the local galaxy sample ALLSMOG (see \citet{Bothwell+2016} for details). The solid curve
 represents the quadratic fit to the ALLSMOG data, here shown using the revised data
 values from \citet{cicone2017}. 
 Our simulations as well as the full Z18 sample are in good agreement with the ALLSMOG results.}
    \label{fig:main_sequence}
\end{figure}

In this section we perform a series of sanity checks of our simulated galaxies by examining their star formation rates, masses (stellar and molecular), and metallicities in the context of known empirical correlations between these quantities. Before doing so, we stress that observed integrated properties such as stellar and molecular gas masses are likely to be uncertain by at least a factor of two \citep{lower2020, gilda2021}.

Fig.\ \ref{fig:main_sequence}a shows the location of our simulated galaxies in the ${\rm SFR}-M_{\rm \star}$ plane, where they define a sequence that extends over more than three orders of magnitude in stellar mass. The three ``clouds'' of simulations correspond with the three volumes of \simba galaxies ($25$, $50$, $100\,{\rm cMpc}\,h^{-1}$; see \S3). The cloud with the highest mass galaxies corresponds with the largest volume, and the cloud with the lowest mass galaxies corresponds with the lowest volume. As explained in \cite{Leung2020}, the galaxies in the larger volumes have a larger smoothing length and lower initial gas mass resolution than in the lowest volume.

Also shown in Fig.~\ref{fig:main_sequence}a, are the ALPINE and $z \geq 4$ Z18 samples.
The galaxy main sequence is usually
modelled by a log-linear relation of the form:  $\log {\rm SFR} = \alpha \log M_{\rm \rm \star} + \beta$, where the parameters $\alpha$ and $\beta$ are functions of cosmic time, or redshift. The normalisation, $\beta$, in particular, is found to increase at earlier cosmic epochs. Observational constraints on the galaxy main sequence at $z\simeq 5-6$ come
from \cite{speagle2014}, \cite{Schreiber2015}, \cite{Salmon2015}, \cite{Pearson2018}, and \cite{Khusanova2020}.
The \cite{speagle2014} main sequence is based on observed galaxy samples at $z\simeq 4-6$ with $M_{\rm \star} = 10^{9.5}-10^{11.5}\,{\rm \Msolar}$, while \cite{Schreiber2015} derived their main sequence from a small sample of galaxies at $z\simeq 3.5-5$ with $M_{\rm \star} \ge 10^{11}\,{\rm \Msolar}$, and an extrapolation to $z\simeq 6$. Finally, the \cite{Salmon2015} main sequence is based on a sample of $z\simeq 6$ galaxies with stellar masses in the range $M_{\rm \star} = 10^{9}-10^{10.5}\,{\rm \Msolar}$. 
 While the main sequence parametrizations by \cite{speagle2014}, \cite{Salmon2015}, and \cite{Khusanova2020} are in agreement with the ALPINE and Z18 data, the parametrizations by \cite{Schreiber2015} and \cite{Pearson2018} appear to have normalizations that fall below that of the data (albeit the slopes appears to match). This may be because the \cite{Schreiber2015} fit was made using $z < 4$ data, and is thus an extrapolation towards higher redshift. Nevertheless, the main sequence parametrizations by \cite{Schreiber2015} and \cite{Pearson2018} are seen to match our simulations across the full stellar mass range spanned by the simulations, although these are based on extrapolations to lower masses than those probed by the observations.

For stellar masses $\geq 2\times 10^9\,{\rm \Msolar}$ there is fair agreement between our simulations, and the ALPINE and $z \geq 4$ Z18 data. At lower stellar masses (down to $\sim 2\times 10^8\,{\rm \Msolar}$) there is an apparently increasing offset between the simulations and the observational data, although, there are few observational data available at these low masses to warrant a fair comparison. We note that both the ALPINE and Z18 data may be somewhat biased toward high SFR-values simply by virtue of having SFR-estimates at these low masses. Furthermore, there may be a selection bias in the ALPINE sources due to the selected parent sample of ALPINE; this is discussed further in \cite{Khusanova2021}.

Fig.~\ref{fig:main_sequence}b shows ${\rm SFR}$ vs $M_{\rm mol}$ for our
simulated galaxies, along with $z\geq 4$ galaxies from our comparison samples.
Our simulated galaxies trace out a correlation which is consistent with the galaxy-integrated version of the Kennicutt-Schmidt (KS) law \citep{Kennicutt1998}. We show parametrizations of the redshift-independent galaxy-integrated KS-law for starbursts (yellow line) and normal disk galaxies (green line) from \citet{Sargent2014}. The galaxies simulated within the $25\,{\rm cMpc}\,h^{-1}$ box are seen to agree with the KS-law for starbursts, while the simulated galaxies extracted from the $50\,{\rm cMpc}\,h^{-1}$ box fall slightly below the starburst KS-law. Galaxies from the $100\,{\rm cMpc}\,h^{-1}$ box-- as well as the observed samples -- fall in between the KS-law for starbursts and disks. They agree well with the Z18 and ALPINE samples, which are seen to also straddle the KS-laws for starbursts and normal disk galaxies.

Finally, in Fig.~\ref{fig:main_sequence}c we examine the relationship between gas-phase metallicity, $M_{\rm \star}$, and $M_{\rm mol}$. \citet{Bothwell+2016} showed that the SFR-metallicity relation at a given stellar mass (the so-called fundamental metallicity relation) is in fact a result of an underlying relation between metallicity and molecular gas mass. Using the ALLSMOG sample of nearby galaxies \citep{Bothwell+2016,cicone2017}, they derived a  relation between metallicity ($12 + \log ({\rm O/H})$), $M_{\rm \star}$, and $M_{\rm mol}$. This relation is shown as the solid line in Fig.\ \ref{fig:main_sequence}c. It represents the linear combination in $\log M_{\rm \star}$ and $\log M_{\rm H_2}$, which yields the least scatter in metallicity (see \citet{Bothwell+2016} for details). 

More recently, however, \citet{cicone2017} reanalysed the ALLSMOG sample and derived improved metallicities, as well as stellar and molecular gas masses. Using these revised and final values, we plot the ALLSMOG sample in Fig.\ \ref{fig:main_sequence}c (orange symbols). 
We find that the galaxies from Z18 are consistent with the ALLSMOG sample, independent of their redshifts, and even extends the relation to the point where it turns over. To overlay the \simba simulations, we use the calibration $12 + \log ({\rm O/H}) = \log ( \langle Z\rangle_{\rm mass}) + 9$ derived by \citet{Ma+2016} using high-resolution cosmological simulations over the redshift range $z=0-6$. \citet{Ma+2016} uses the mass-weighted average gas-phase metallicity, $\langle Z\rangle_{\rm mass}$, and for that reason we have used this quantity from our simulations, and not the SFR-weighted average gas-phase metallicities (see \S\ref{subsection:simulations}), to infer $12+\log ({\rm O/H})$. Our simulations are seen to be in good agreement with the observations in the region where there is overlap, and they show a declining trend towards lower metallicities. 


The analysis above demonstrates that within the range of masses where a comparison with observations is possible, our simulated \simba galaxies are in overall agreement with key scaling relations observed for $z \geq 4$ main sequence galaxies. This suggest that the simulations are representative, even at the low stellar mass-end that extends well below the observed range, of the bulk of the normal star forming galaxies at these epochs.

\subsection{ISM properties}\label{subsection:ISM-properties}
Before we explore the H$_2$-tracing capabilities of [C{\sc ii}], it is instructive 
to first examine the gas mass fractions of the three ISM phases in our simulations 
and their [C{\sc ii}] emission properties. We note here that the diffuse, molecular, and atomic gas masses are calculated to be the total mass of all gas particles associated with each respective phase according to \caesar{}. However, for observations these phases are typically calculated within a certain radius. Therefore the gas mass fractions are calculated to be independent of galactic region or radius for our simulations.

\begin{figure}
 \centering
 \includegraphics[width=\columnwidth]{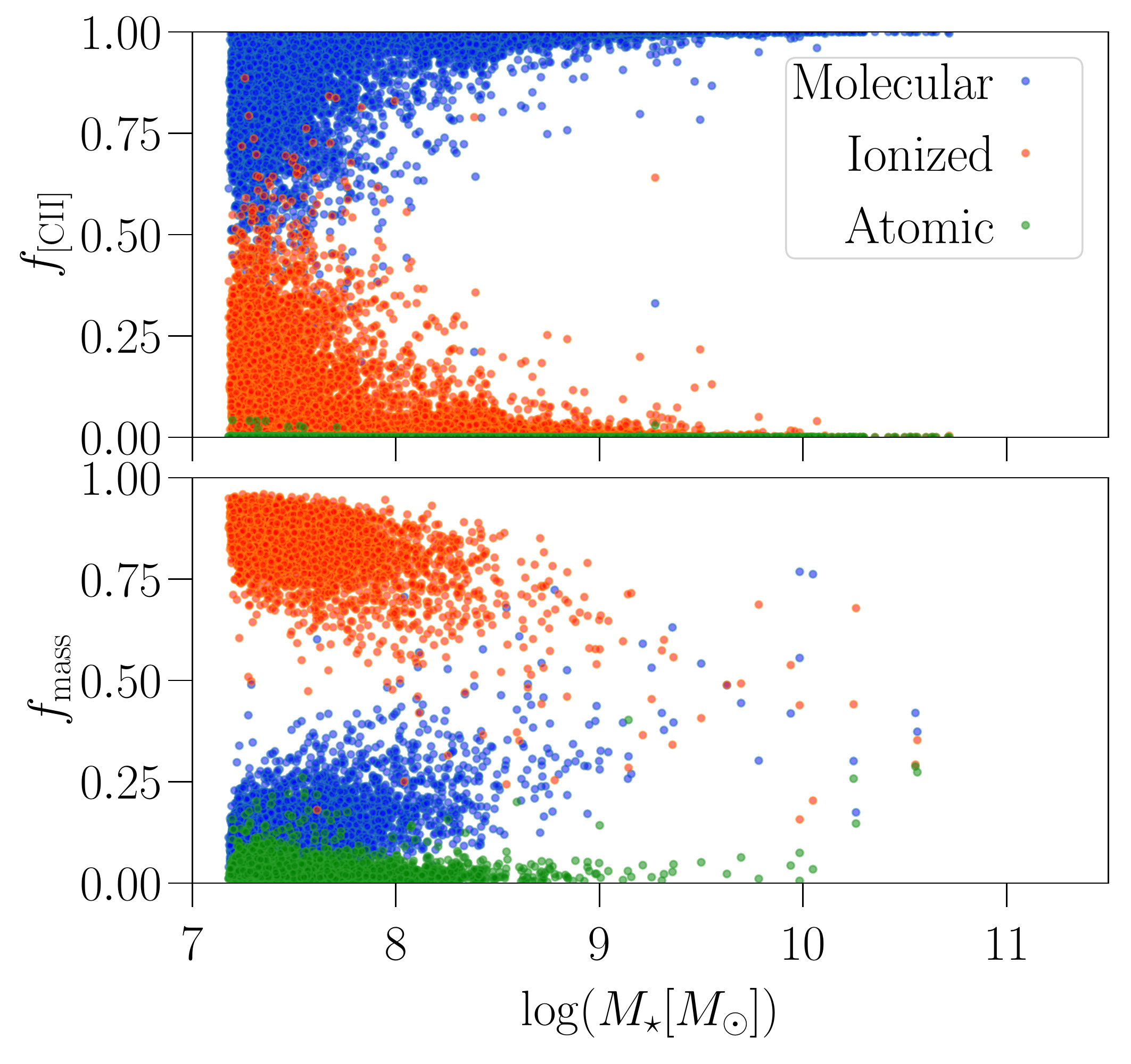}
     \caption{{\bf Top:} The fraction of the total \cii luminosity emitted by the molecular (blue), ionised (red), and atomic (green) ISM phases for each galaxy as a function of its stellar mass. {\bf Bottom:} the same as in the top panel but now showing the gas mass fraction of the different ISM phases.}
     \label{figure:ism-phases}
 \end{figure}

Fig.~\ref{figure:ism-phases} shows fractions of total \cii luminosity ($f_{\rm \cii}$) and ISM gas mass ($f_{\rm mass}$) as a function of stellar mass. It is seen that the molecular phase dominates the \cii emission: it contributes $\sim 50\%$ to 
the total \cii emission for galaxies with $M_{\rm \star} \sim 10^7\,{\rm \Msolar}$ and $\sim 100\%$ for $M_{\rm \star} \gs 10^9\,{\rm \Msolar}$ galaxies. Over the same stellar mass range, the ionized gas phase contribution to the
total \cii emission spans $\sim 50\%$ at the low-mass-end to only a few per-cent at the high-mass-end.
The atomic gas hardly contributes anything to the \cii emission. Looking at the gas mass fraction, the ionized and molecular gas
phases dominate, with the former accounting for $\sim 50-100\%$ and the latter with $\sim 10-75\%$.  Thus, while the molecular gas does not account for most of the ISM mass in our simulated galaxies, except for 
in the most massive galaxies ($M_{\rm star}\gs 10^{9.5}\,{\rm \Msolar}$), it is responsible for the bulk of the
total \cii emission. 

In contrast, the atomic phase makes up $\ls 10\%$ of the total ISM mass, and even less of the total \cii emission. The low mass fraction for the atomic gas is somewhat surprising, given that 
the cosmic atomic gas density exceeds that of molecular gas at redshifts $z\gs 3$ \citep{Walter2020}. We believe that this is due to
the way that gas is classified as 'ionized' in \sigame, which is simply when the electron-to-hydrogen fraction is $x_{e^-} > 0.5$. As a result, some of the atomic gas is lumped in with the ionized gas, and both the true atomic gas mass and \cii emission fraction are thus likely to be higher than what is
indicated in Fig.~\ref{figure:ism-phases}. 

The fraction of the \cii emission coming from the molecular phase in our simulations is consistent with what
is found in local star-forming galaxies \citep[$\sim 60-80\%$;][]{Accurso2017}, as well as in high-resolution simulations
of $z\simeq 6$ Lyman-break galaxies \citep[$\sim 95\%$;][]{pallottini2017}. The contribution to the \cii
emission from the ionized phase is seen to increase as the stellar
mass, and thus the metallicity, decreases. Although, one would expect the \cii cooling rate to decrease at lower metallicity, this effect is negligible compared to the increase in CO photo-dissociation rate, and thus available C$^+$ ions, that comes with lower metallicities; a similar effect was seen by \cite{Accurso2017}. The fact that the bulk of the \cii emission is coming from the molecular gas suggests that we can indeed use the line as a molecular gas tracer. 

\section{Results and Discussion}

\subsection{How well does \cii trace molecular gas mass?}\label{section:mgas-cii}
In this section we examine whether there is a correlation between the [C{\sc ii}] emission and the molecular gas masses of 
our simulated $z\simeq 6$ galaxies, and to what extent this correlation agrees with the empirical relation found by observations.
 \begin{figure}
 \centering
 \includegraphics[width=\columnwidth]{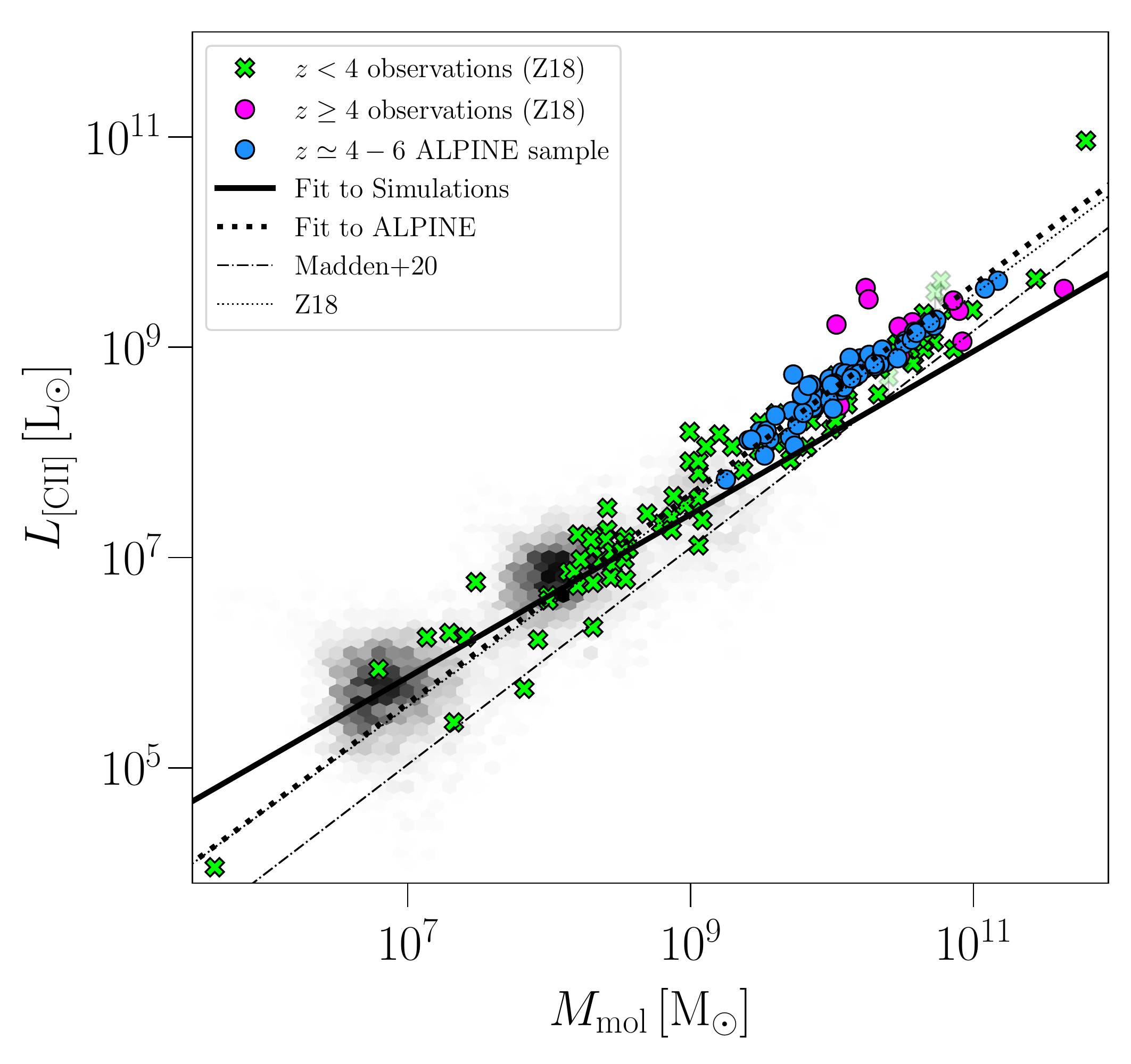}
  \includegraphics[width=\columnwidth]{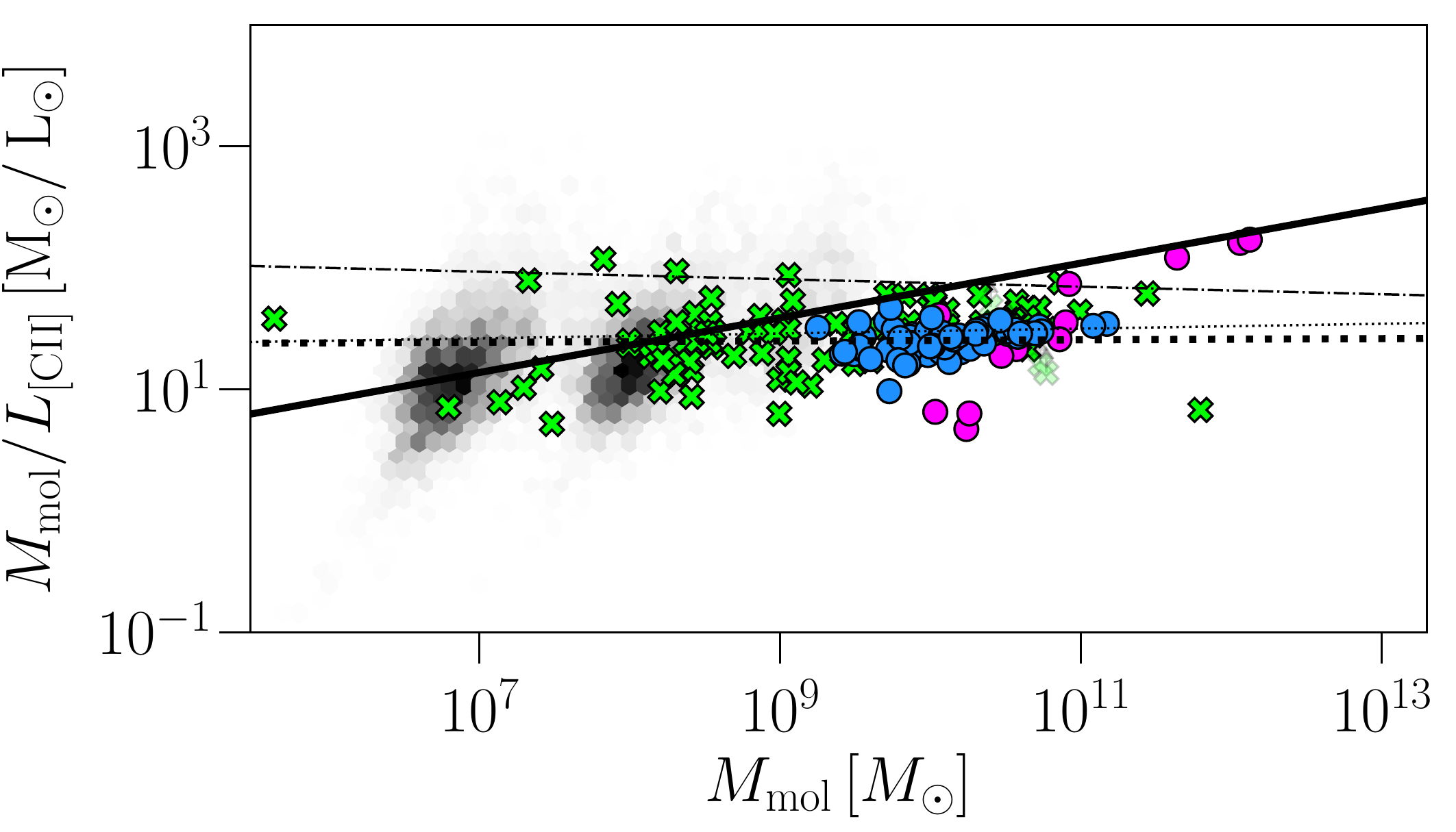}
     \caption{ $L_{\rm [CII]}$ vs $M_{\rm mol}$ (top) and $M_{\rm mol}/L_{\rm [CII]}$ vs $M_{\rm mol}$ (bottom) for our simulated $z=6$ galaxies (grey hexbin) and observed comparison samples from Z18 and \citet{dessauges2020}. Log-linear fits to our simulations (solid line), the observed samples (dotted lined), and to simulations by \citet{madden2020} of nearby20warf galaxies (dot-dashed line) are indicated. } \label{figure:mgas-cii}
 \end{figure}

Fig.~\ref{figure:mgas-cii} shows the [C{\sc ii}] luminosity vs the total molecular gas mass and $M_{\rm mol}/L_{\rm \cii}$ vs $M_{\rm mol}$ for our simulated galaxies, along with observed $z < 4$ and $z \geq 4$ galaxies from Z18, and $z = 4-6$ galaxies from the ALPINE survey \citep{dessauges2020}. We fit a log-linear model of the form $\log L_{\rm \cii}
= a \log M_{\rm mol} + b$ to the simulations only, as well as the ALPINE data only. This yields the following relations:

\bigskip

\noindent Simulations:
\begin{eqnarray}
    \log L_{\rm \cii} &=& (0.78\pm 0.01) \log M_{\rm mol} +
    (0.44\pm 0.04),
    \label{equation:mgas_cii-sims}
\end{eqnarray}
shown as the thick solid line in Fig.~\ref{figure:mgas-cii}. The r.m.s.~scatter of the simulations around the fitted line is $0.45\,{\rm dex}$.

\bigskip

\noindent ALPINE:
\begin{eqnarray}
    \log L_{\rm \cii} &=& (1.00\pm 0.01) \log M_{\rm mol} -
    (1.35\pm 0.1),
    \label{equation:mgas_cii-alpine}
\end{eqnarray}
shown as the thick dashed line in Fig.~\ref{figure:mgas-cii}, with an r.m.s.~scatter of the data around the fitted line
of $0.12\,{\rm dex}$. 

For comparison, Z18 derived a relation given by 
$ \log L_{\rm \cii} = (0.98\pm 0.02) \log M_{\rm mol} - (1.28\pm 0.01)$,
which is shown as the thin dotted line in Fig.~\ref{figure:mgas-cii}. They reported an r.m.s.~scatter of $0.2\,{\rm dex}$. 
There is good agreement between the observed data from ALPINE and Z18 \citep[see also ][]{dessauges2020}, with the fit to the ALPINE data being very similar to the Z18-relation. The Z18 data exhibit somewhat larger scatter compared to the ALPINE data.  This is likely due to the ALPINE galaxies being uniformly selected as main-sequence galaxies at $z=4-6$, while the Z18 sample is heterogeneous, consisting of local (U)LIRGs, as well as main-sequence galaxies and starburst at low and high redshifts.

Our simulations appear to be in broad agreement with the observations, despite lying in a very different part of the 'galaxy parameter space' (see \S\ref{section:sample-properties}).
Upon closer inspection, it is seen from Fig.~\ref{figure:mgas-cii} that our simulations contain a number of galaxies which exhibit higher \cii luminosities given their gas masses than would be expected from the Z18 relation. On average, the simulations have lower $M_{\rm mol}/L_{\rm [CII]}$ values than bulk of the observed data. This is most pronounced for the lowest gas mass simulations ($M_{\rm mol} \ls 10^7\,{\rm M_{\odot}}$), while the simulations with the highest gas masses ($M_{\rm mol} \sim 10^9\,{\rm M_{\odot}}$) have $M_{\rm mol}/L_{\rm [CII]}$ ratios in line with the observed data. The overall effect of this is a shallower slope in the simulations only relation, and a higher r.m.s.~scatter ($0.45\,{\rm dex}$).
The shallower slope ($\sim 0.78$) suggested by our simulations implies a $L_{\rm \cii}$-dependent, i.e., non-universal $M_{\rm mol}/L_{\rm \cii}$ ratio for $z\sim 6$ galaxies (cf.~Z18). The small number of \cii observations of galaxies with gas masses $\ls 10^8\,{\rm M_{\odot}}$ prevents us from assessing whether a similar trend is seen in observations.

From the $z\sim 0$ simulations presented by \citet{madden2020}, they derive: $\log L_{\rm \cii} = 1.03 \log M_{\rm mol} - 2.19$, with an r.m.s.~scatter of $0.14\,{\rm dex}$. This relation, shown as the thin dash-dotted line in Fig.~\ref{figure:mgas-cii}, has a slope consistent with unity, i.e., similar to the Z18 and ALPINE relations, but the normalisation is significantly lower. Using this relation would result in $\sim 3\times$ larger molecular gas masses, which they ascribe to Z18 not accounting for CO-dark gas and thus using a CO-to-H$_2$ conversion factor that is too low \citep{madden2020}. This explanation, however, is not borne out by our simulations, which examine the relationship between $L_{\rm \cii}$ and the full molecular gas reservoir (CO-dark or not).                

\subsection{How well does \cii trace star formation rate?}\label{section:SFR-cii}

\begin{figure}
 \centering
 \includegraphics[width=\columnwidth]{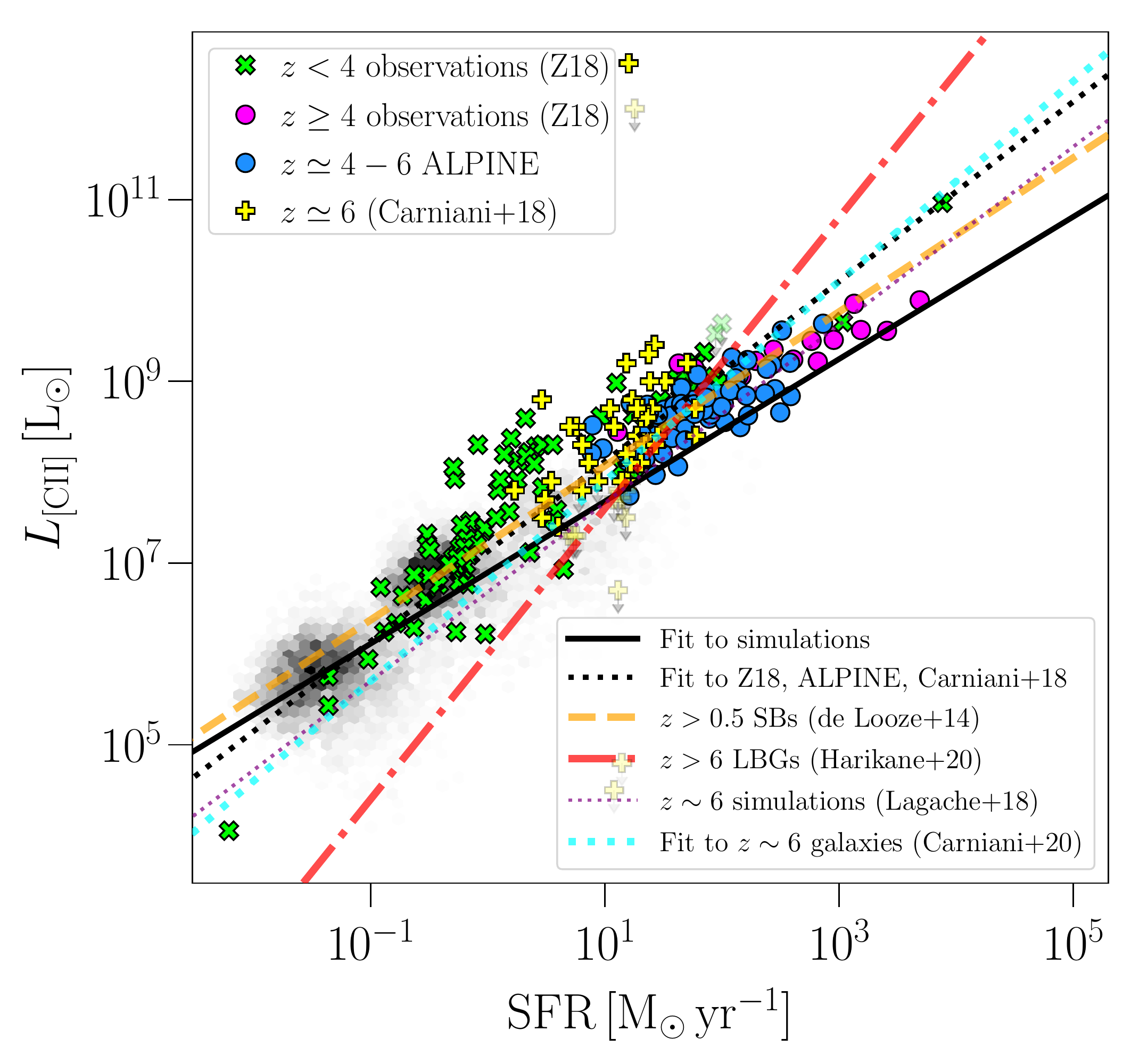}
  \includegraphics[width=\columnwidth]{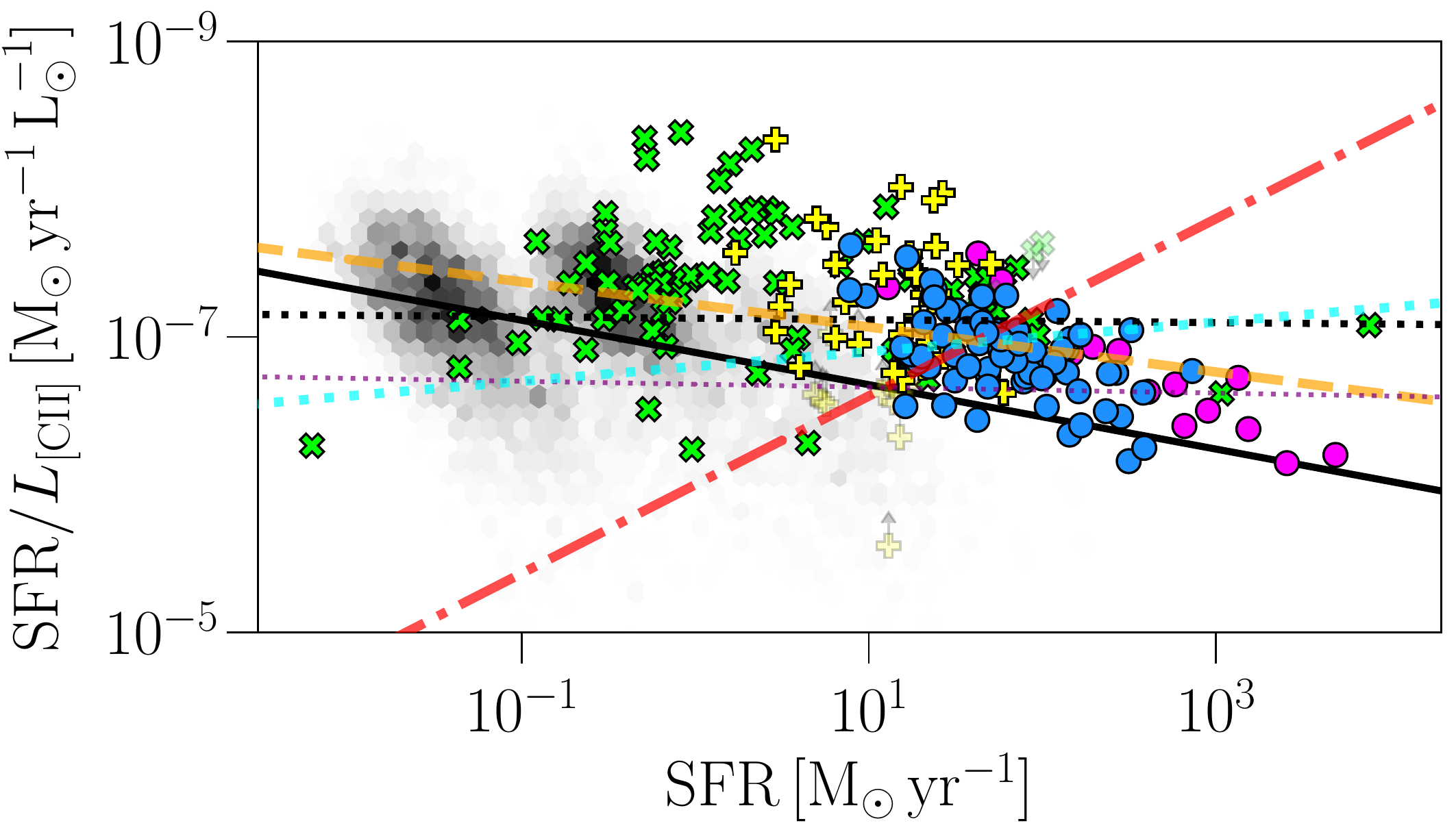}
     \caption{$L_{\rm [CII]}$ vs. SFR (top) and SFR/$L_{\rm [CII]}$ vs. SFR (bottom) for our simulated $z=6$ galaxies (grey hexbin), and the observed comparison samples from Z18, \citet{dessauges2020}, and \citet{Carniani2018} (circles and crosses).} \label{figure:sfr-cii}
 \end{figure}

The traditional interpretation of the \cii line has been as a diagnostic of the physical conditions in photon-dominated regions, and as a tracer of the star formation activity in galaxies \citep{Stacey1991, delooze2014}. In Fig.~\ref{figure:sfr-cii} we show $L_{\rm \cii}$ vs ${\rm SFR}$ for our simulations, the ALPINE, the Z18 sample, and the \citet{Carniani2018} (henceforth C18) sample.
Also shown is the log-linear fit to the combined Z18, C18, and ALPINE samples, as well as our simulations.

For comparison the global $L_{\rm \cii}-{\rm SFR}$ relation inferred for high-$z$ starbursts by \citet{delooze2014} is also shown, along with relations derived from observations \citep{harikane2020} and simulations \citep{lagache2018} of $z\simeq 6$ star-forming galaxies.
The r.m.s.~scatter  of  the  $\log L_{\rm [CII]}-\log {\rm SFR}$ relation fitted to our simulations alone is $0.50\,{\rm dex}$. A fit to the Z18, ALPINE, and C18 data exhibits an even larger scatter, i.e., $0.55\,{\rm dex}$, which is higher than the scatter reported by \citet{delooze2014} for high-$z$ starbursts ($0.40\,{\rm dex}$). This is also higher than \citet{schaerer2020}, who found a scatter of 0.28 dex when plotting $L_{\rm [CII]}$ vs. SFR of the ALPINE sample after using the IRX-$\beta$ relation \citep[][]{fudamoto2020} to add an $L_{\rm IR}$-derived SFR to the UV-derived SFRs \citep[][also see \S 2]{faisst2020}.

The $\log L_{\rm \cii} - \log {\rm SFR}$ relation exhibits more scatter than the $\log M_{\rm mol} - \log L_{\rm \cii}$ relation, both for our simulations and for observed data, and we thus conclude that the \cii luminosity is more tightly correlated with the molecular gas mass than the star-formation rate. Our simulations are seen to agree well with the high-$z$ relation derived by \citet{delooze2014}, which also predict a slightly decreasing ${\rm SFR}/L_{\rm \cii}$ with increasing ${\rm SFR}$. This trend can account for some of the scatter in the $\log L_{\rm \cii} - \log {\rm SFR}$. The scatter can also arise from other factors; we note, for instance, that SFR is prone to large observational systematics -- particularly at high-redshift -- and this may be the reason why the scatter in the  $\log L_{\rm \cii}- \log {\rm SFR}$ relation is larger than the $\log M_{\rm mol}-\log L_{\rm \cii}$ relation. Furthermore, the methods of measuring SFR, as well as uncertainties on SFR measurements, can affect the total scatter on the $\log L_{\rm [CII]}-\log {\rm SFR}$ relation for our observations.

\begin{figure}
    \centering
        \includegraphics[width=\columnwidth]{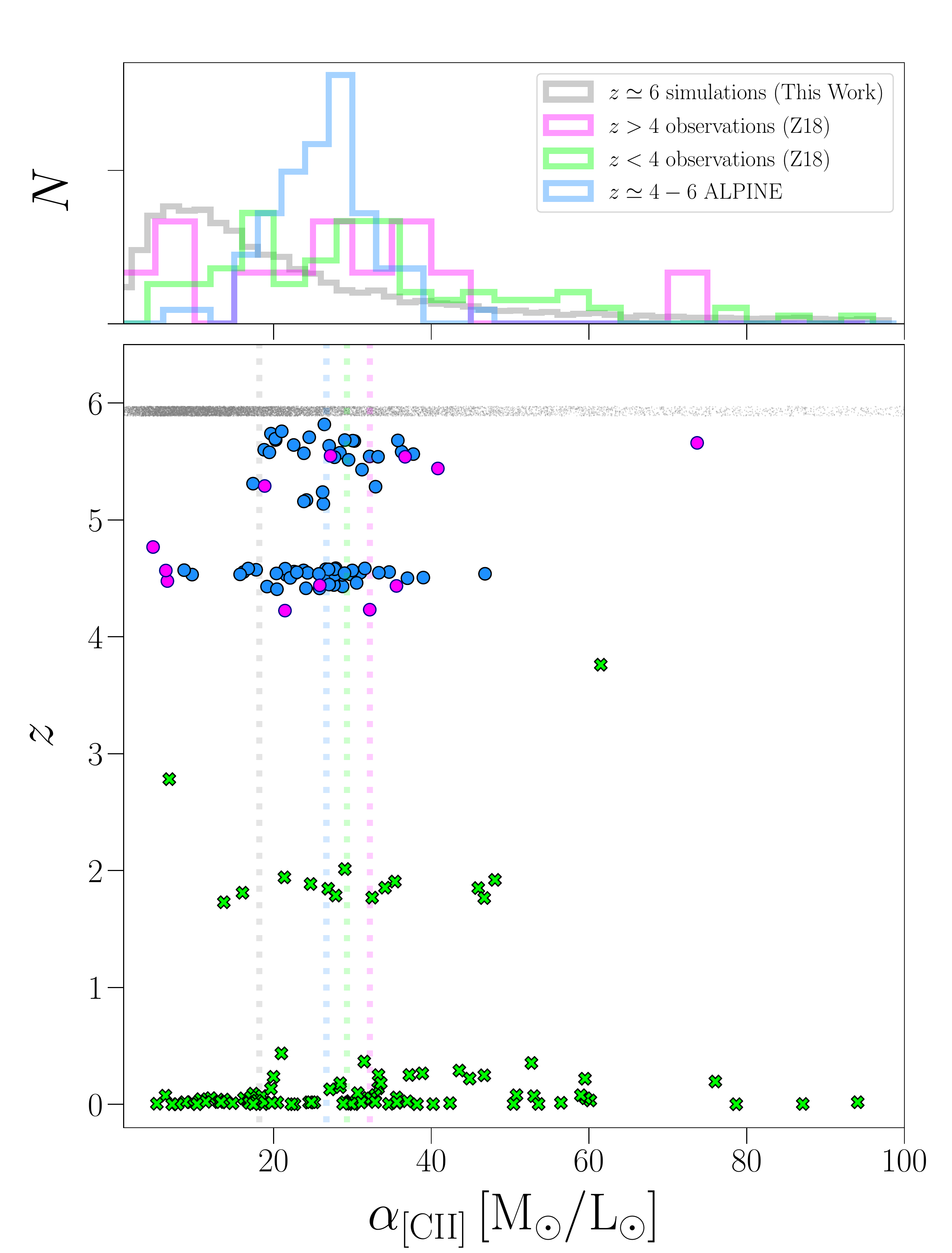}
    \caption{{\bf Top panel:} The distributions of $\alpha_{\rm \cii}$ values for our simulations (grey horizontal band) and comparison samples. {\bf Bottom panel:} $\alpha_{\rm \cii}$ vs redshift for our simulations and comparison samples. The vertical dashed lines indicate the median $\alpha_{\rm [CII]}$ values of the different samples. For clarity, our $z=6$ simulations have been spread out in the $z$-direction slightly.}
    \label{fig:alpha_cii_vs_z}
\end{figure}{}

\begin{figure}
    \centering
    \includegraphics[width=\columnwidth]{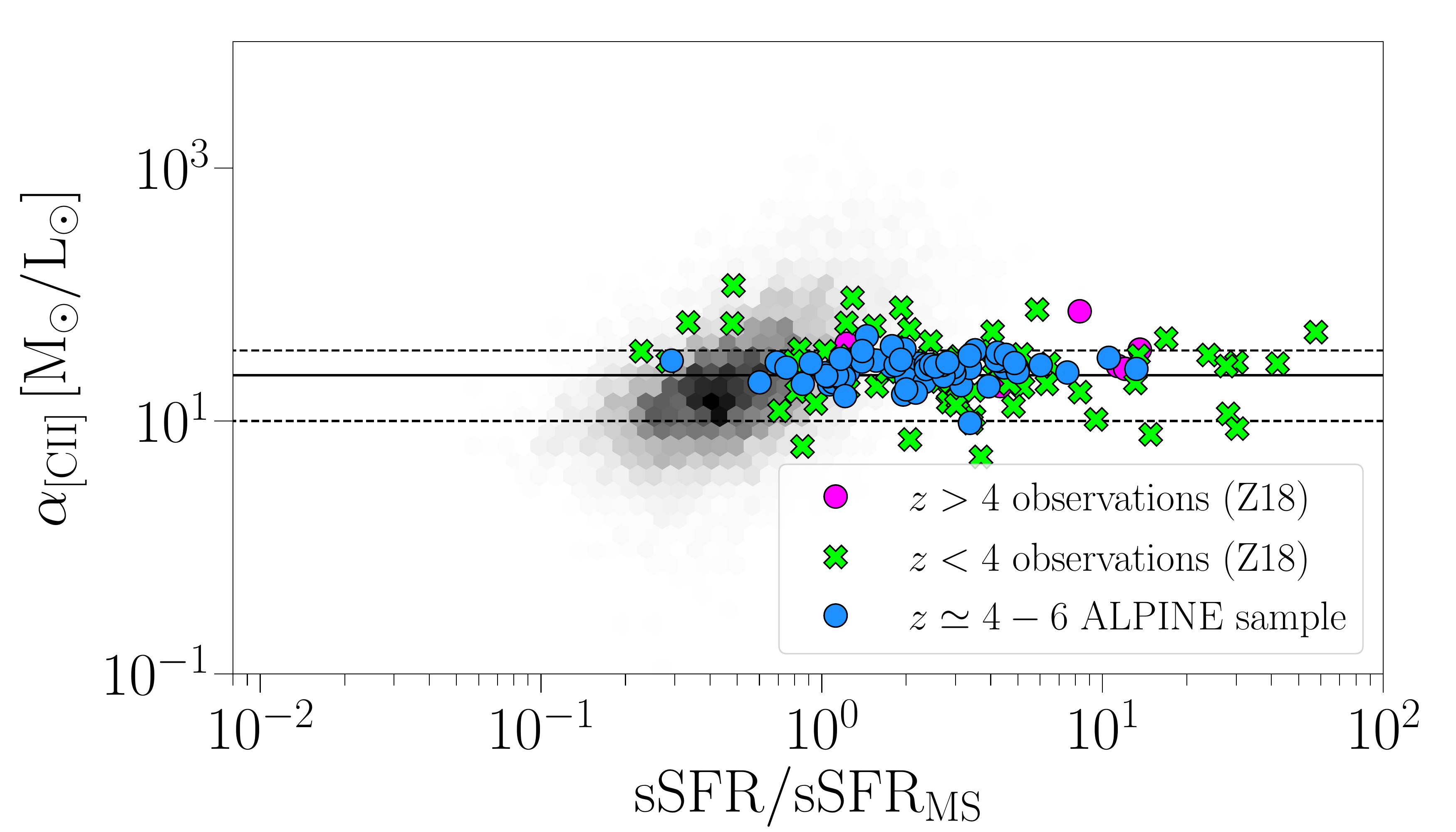}
    \caption{$\alpha_{\rm [CII]}$ vs offset from the main sequence for our simulations (grey hexbins) and observed galaxies from Z18 (green and magenta symbols), and the ALPINE survey \citep[blue filled circles,][]{dessauges2020}.} 
    \label{figure:alpha_cii_dms}
\end{figure}{}

\subsection{The \cii conversion factor}\label{subsection:alpha-cii}
The fact that Z18 found the $L_{\rm \cii}-M_{\rm mol}$ relation to be log-linear with a slope close to unity allowed them to define a [C{\sc ii}] luminosity to molecular gas mass conversion factor, $\alpha_{\rm \cii} = M_{\rm mol}/L_{\rm \cii}$. They inferred a value of $\alpha_{\rm \cii} \sim 30\,{\rm \Msolar/\Lsolar}$ with an uncertainty of $0.3\,{\rm dex}$ averaged across their sample, and argued that $\alpha_{\rm \cii}$ was effectively invariant with redshift, metallicity, and galaxy type (e.g., starburst vs main sequence galaxy), see also Fig.~\ref{figure:mgas-cii}. 
For the ALPINE sample, we find a median conversion factor of $\alpha_{\rm \cii}\sim 27\,{\rm \Msolar/\Lsolar}$ and a median absolute deviation (MAD) of $4\,\rm M_{\sun}/L_{\sun}$. 
From our simulated galaxies, we find a median value of $\alpha_{\rm \cii}\sim 18\,{\rm \Msolar/\Lsolar}$, with a MAD of $10\,\rm M_{\sun}/L_{\sun}$. 

In Fig.\,\ref{fig:alpha_cii_vs_z} we plot $\alpha_{\rm \cii}$ vs $z$ for our simulations, the Z18 and ALPINE samples along with  histograms of their overall $\alpha_{\rm \cii}$ distributions (top panel). The $\alpha_{\rm \cii}$-distribution for our simulated galaxies peaks at $15\,\rm M_{\sun}/L_{\sun}$ with an extended tail towards higher values.
The $\alpha_{\rm \cii}$-distribution for the Z18 sub-sample at $z < 4$ also exhibit a tail towards high $\alpha_{\rm \cii}$-values. In addition to high $\alpha_{\rm \cii}$-values, the simulated sample also contain many galaxies with $\alpha_{\rm \cii}$-values $< 10\,\rm M_{\sun}/L_{\sun}$. These low values of $\alpha_{\rm \cii}$ derived from our simulations are in agreement with high-$z$ \cii observations by \cite{rizzo2021}, who inferred a median $\alpha_{\rm \cii} = 7^{+4}_{-1}$ \msun/\lsun for five strongly-lensed starburst galaxies between $4 \leq z \leq 5$ \citep[see also][]{}{rizzo2020}.
Finally, we note that the Z18 sub-sample at $z \geq 4$ overlaps with the $\alpha_{\rm \cii}$-values of the ALPINE sample, and both are tightly distributed around their peak. For the ALPINE sample, in particular, this might be due to it being a relatively homogeneous sample of main sequence galaxies. In contrast, the full Z18 sample, as well as our simulations span a wider range in star formation rate and masses.

 In Fig.\,\ref{figure:alpha_cii_dms} we show $\alpha_{\rm \cii}$ vs offset from the main sequence for our simulated galaxies, which for a given galaxy is defined as the ratio of its specific star formation rate (${\rm sSFR} = {\rm SFR}/M_{\rm \star}$) and the specific star formation of a galaxy with the same stellar mass on the main sequence relation (at the given redshift). We adopt the prescription for the main sequence at $z=6$ as given by \citet{Schreiber2015}, since it agrees well with our simulations (\S\ref{subsection:integrated-properties}). The Z18 and ALPINE comparison samples have also been normalized with the \citet{Schreiber2015} main-sequence parametrisations at the relevant redshifts of each individual galaxy. Our simulations show a weak trend of increasing $\alpha_{\rm [CII]}$ with main-sequence offset.  
 In contrast, both the Z18 and ALPINE samples show constant $\alpha_{\rm [CII]}$. This is true even for the sources with large offsets from the main-sequence.

\subsection{Is [C{\sc ii}] a metallicity-invariant tracer of molecular gas?}\label{subsection:invariance}
\begin{figure}
    \centering
    \includegraphics[width=\columnwidth]{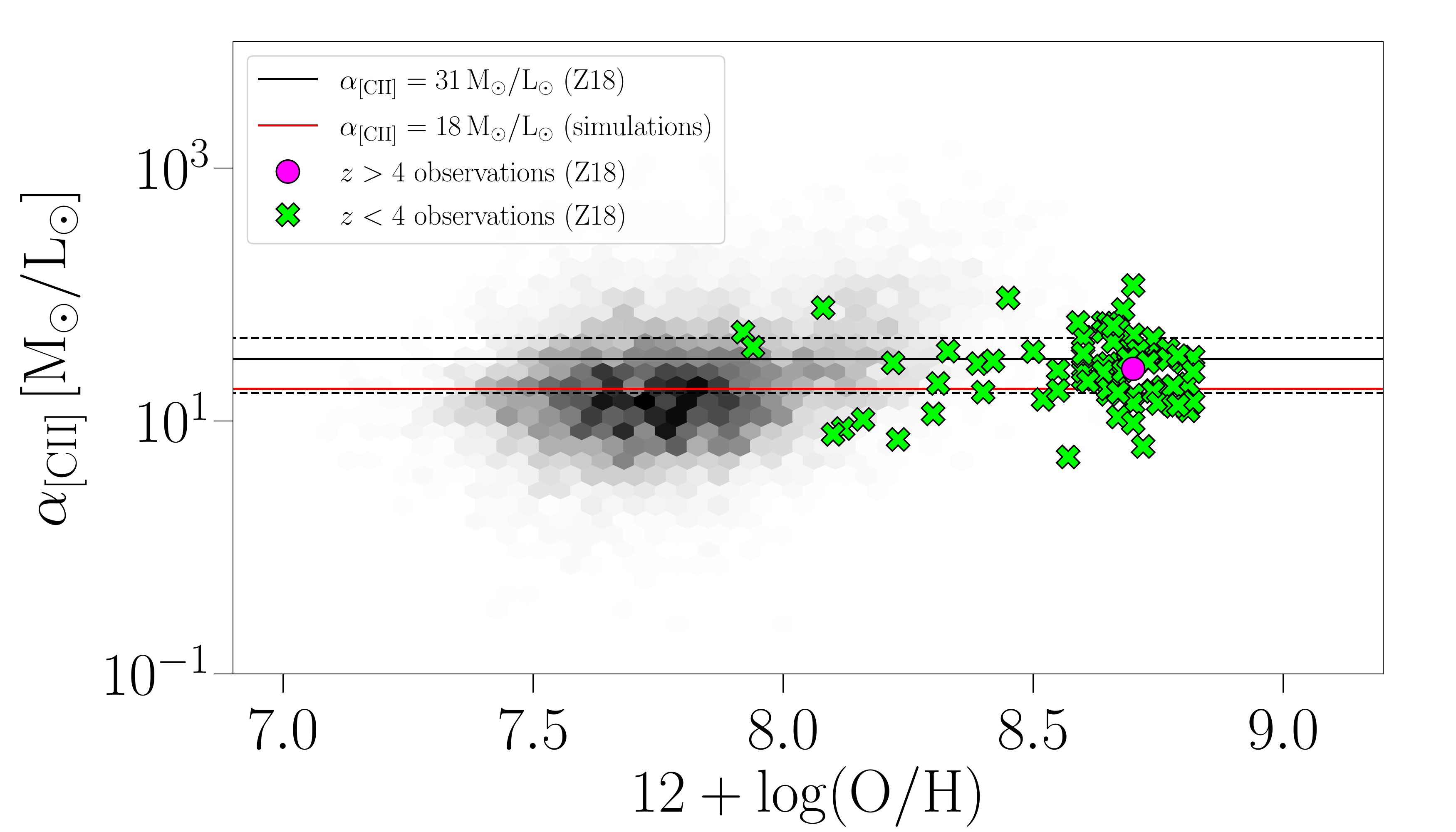}
    \includegraphics[width=\columnwidth]{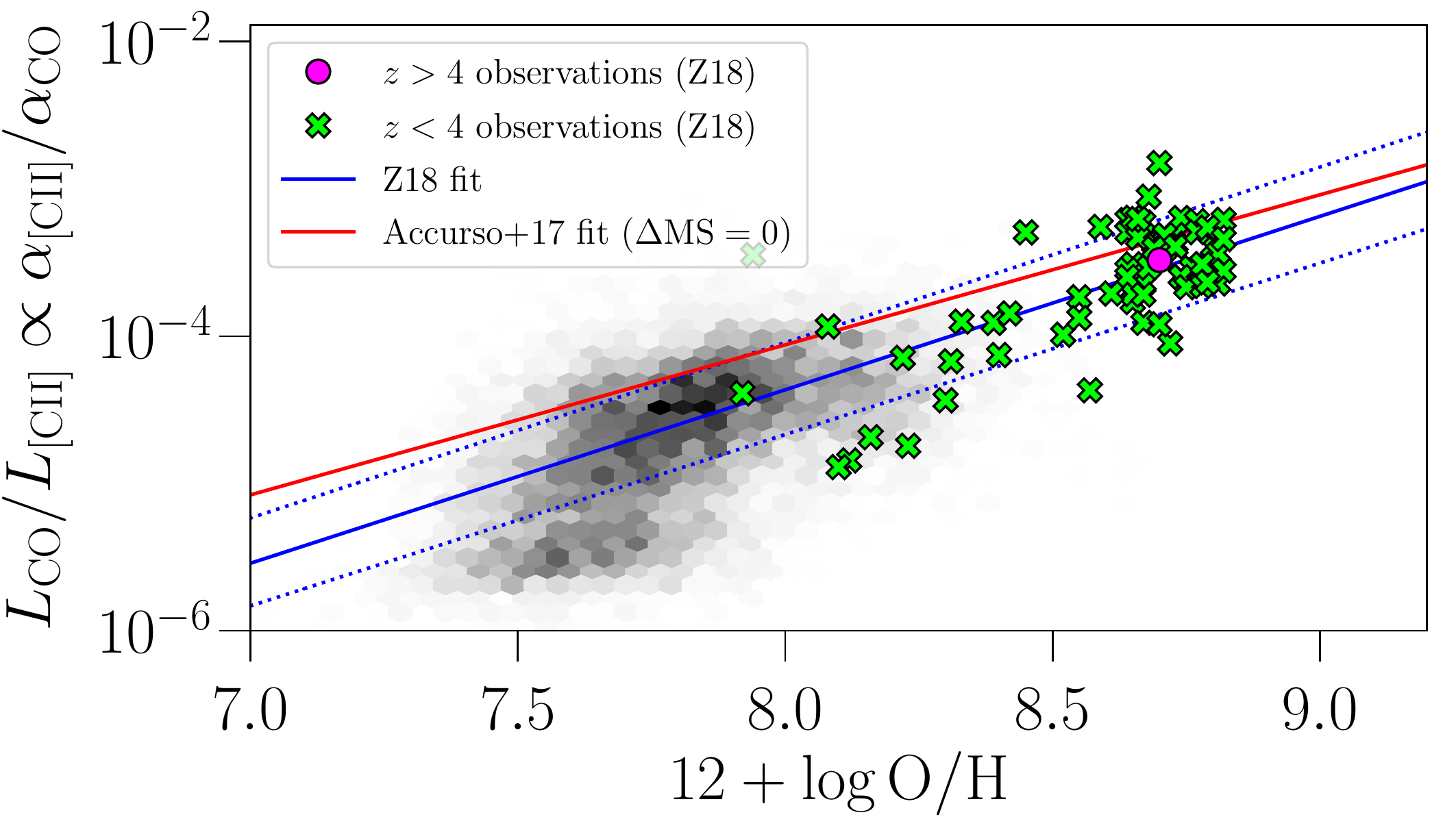}
    \caption{{\bf Top:} $\alpha_{\rm [CII]}$ vs gas-phase metallicity for our simulated galaxies (gray hexbins), along with observed samples from Z18 (green and magenta symbols). The black solid line indicates the median $\alpha_{\rm [CII]}$ value ($31\,{\rm M_{\odot}/L_{\odot}}$) found by Z18 and its $0.2\,{\rm dex}$ scatter (dotted lines), while the red line indicates the median $\alpha_{\rm [CII]}$ value ($18\,{\rm M_{\odot}/L_{\odot}}$) of our $z=6$ simulations. {\bf Bottom:} The CO(1-0)-to-\cii luminosity ratio (both in units of solar luminosities) vs the gas-phase metallicity. The luminosity ratio, which scales with the ratio of the \cii-to-CO(1-0) conversion factors ($\alpha_{\rm \cii}/\alpha_{\rm CO}$), is seen to strongly decrease with decreasing metallicity. This is a result of the increasing photo-dissociation of CO in unshielded environments in metal-poor galaxies. The blue and red lines are fits from \citep{zanella2018} and \citep{Accurso2017}, respectively. } \label{figure:alpha_cii_vs_metallicity}
\end{figure}

In Fig.\,\ref{figure:alpha_cii_vs_metallicity} (top) we show $\alpha_{\rm \cii}$ vs gas-phase metallicity (i.e., $12 + \log({\rm O/H})$) for our simulations along with the  Z18 sample (note, we do not have metallicity estimates for the ALPINE sample). Across the metallicity range probed by our simulations ($12+\log {\rm O/H} \sim 7.0$ to $\sim 9.0$), the $\alpha_{\rm \cii}$ values show no strong dependence on metallicity. As already noted, the simulated $\alpha_{\rm \cii}$ values are, on average, lower than observed values, although with significant overlap between the two distributions. Only few observations, however, are of galaxies with as low metallicities as the simulations, and where there is overlap in terms of metallicity, the agreement is good between the observed and simulated $\alpha_{\rm \cii}$ values (although, the data points are sufficiently sparse that a quantitative comparison is difficult). 
Fig.~\ref{figure:alpha_cii_vs_metallicity} suggests that the on average lower $\alpha_{\rm [CII]}$-value for the simulations is, at least in part, due to their much lower metallicities.
Z18 found $\alpha_{\rm [CII]}$ to be invariant with the metallicity. It should be noted, however, that bulk of their sample fall within a relatively small range of metallicities, making it hard to tease out any trend. Our simulations suggest that there is in fact a weak dependency on metallicity, that becomes apparent when one probes extremely low metallcities
($12 + \log ({\rm O/H}) \ls 8.0$), i.e., much lower metallicities
than probed by Z18, and as low as those of local metal-poor dwarfs \cite{delooze2014}. 
 
Naively, one would expect $L_{\rm \cii}$ to increase with metallicity, due to a higher abundance of carbon. 
This in turn would imply a decreasing trend in $\alpha_{\rm \cii}$ with metallicity, which is not what is seen in  
Fig.\,\ref{figure:alpha_cii_vs_metallicity}(top). However, an increase in metallicity also implies a higher dust content; thus, fewer UV-photons are
available to ionize neutral carbon  (photoionization potential of $11.3 \,{\rm eV}$). It also implies fewer UV-photons capable of photodissociating CO (requires photon energies $> 11.1\,{\rm eV}$) in neutral PDRs, which ultimately means less carbon available for ionization. A higher dust content also implies more UV-shielded regions,
which would give rise to an increase in $\alpha_{\rm [CII]}$ with
gas-phase metallicity.

In \sigame, the interstellar radiation field that impinges on the diffuse and molecular clouds is attenuated by the dust inside the clouds. The fraction of ionized carbon thus depends on the radiation field hitting the clouds, as well as the dust fraction (which scales with the metallicity) of the clouds.
CO as a tracer of molecular gas is known to deteriorate in low-metallicity environments, where the
photo-dissociation of CO leads to a significant fraction of the molecular ISM being CO-dark, \citep[e.g.][]{narayanan2017, li2018, madden2020}. 

This effect is illustrated in Fig.~\ref{figure:alpha_cii_vs_metallicity} (bottom), where the CO-to-[CII] luminosity ratio ($L_{\rm CO}/L_{\rm [CII]} \propto \alpha_{\rm [CII]}/\alpha_{\rm CO}$) is seen to sharply decrease with decreasing metallicity. This is a direct result of $\alpha_{\rm CO}$ increasing with decreasing metallicity (e.g., CO-dark gas), while $\alpha_{\rm [CII]}$ remains largely invariant as is seen in Fig.~\ref{figure:alpha_cii_vs_metallicity} (top). Across the metallicity range $7.0 \ls 12 + \log ({\rm O/H}) \ls 9.0$, $\alpha_{\rm [CII]}/\alpha_{\rm CO}$ increases by more than two orders of magnitude, which is in line with a similar trend found by Z18 and \citet{Accurso2017} (red and blue lines in Fig.~\ref{figure:alpha_cii_vs_metallicity} (bottom)).
 

\subsection{What is the origin of the $\log M_{\rm mol} - \log L_{\rm [CII]}$ relation?}\label{section: digging deeper}
In this section we examine whether combining the log-linear relations $\log L_{\rm [CII]} - \log M_{\rm mol}$ and $\log L_{\rm [CII]} - \log {\rm SFR}$, derived for our simulated galaxies in the previous section, 
leads to a better determination of $M_{\rm mol}$.
Both $L_{\rm [CII]}$ and ${\rm SFR}$ are observables that can be measured
with some confidence towards $z\sim 6$ galaxies.
 \begin{figure}
 \centering
 \includegraphics[width=\columnwidth]{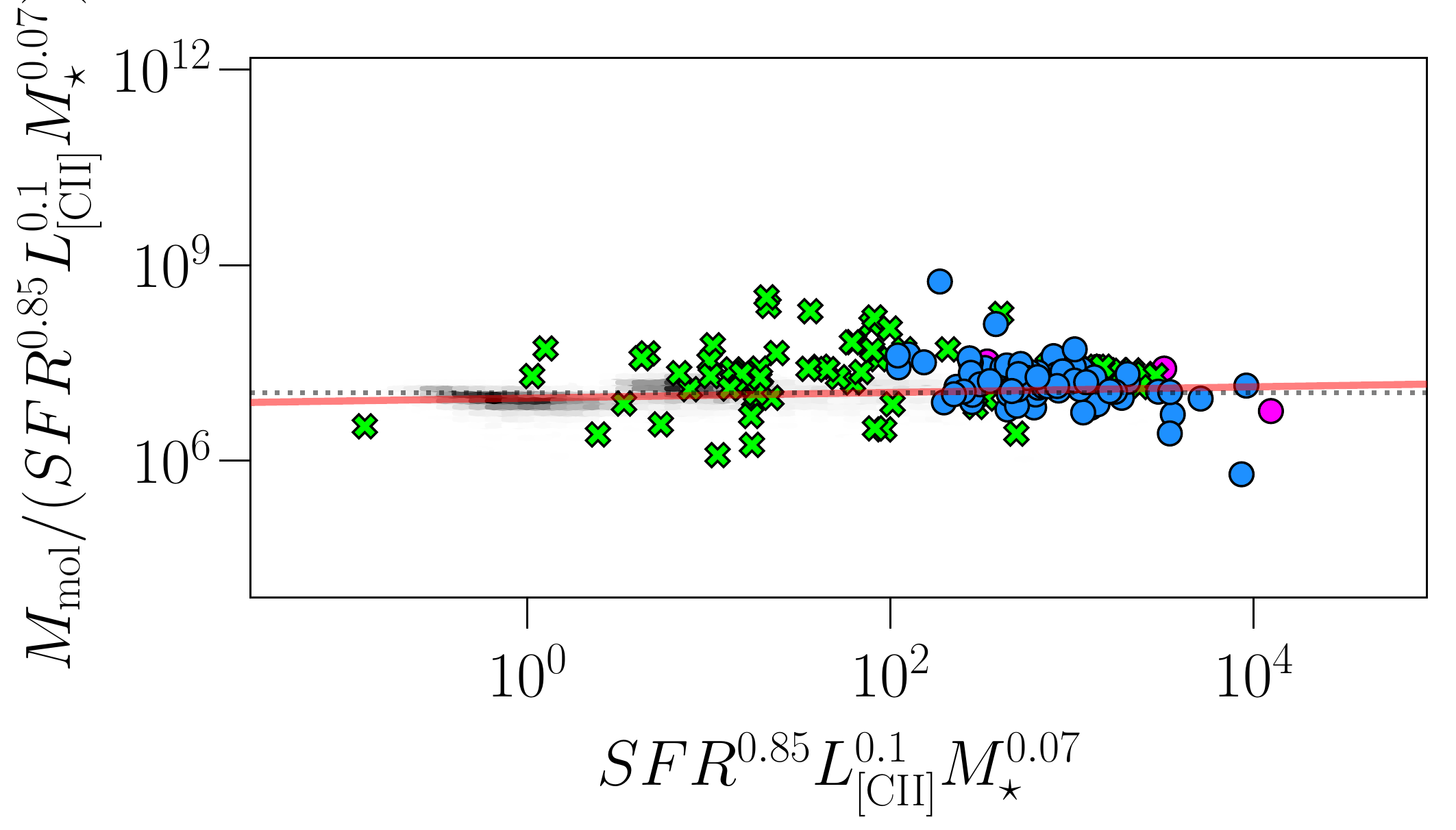}
 \includegraphics[width=\columnwidth]{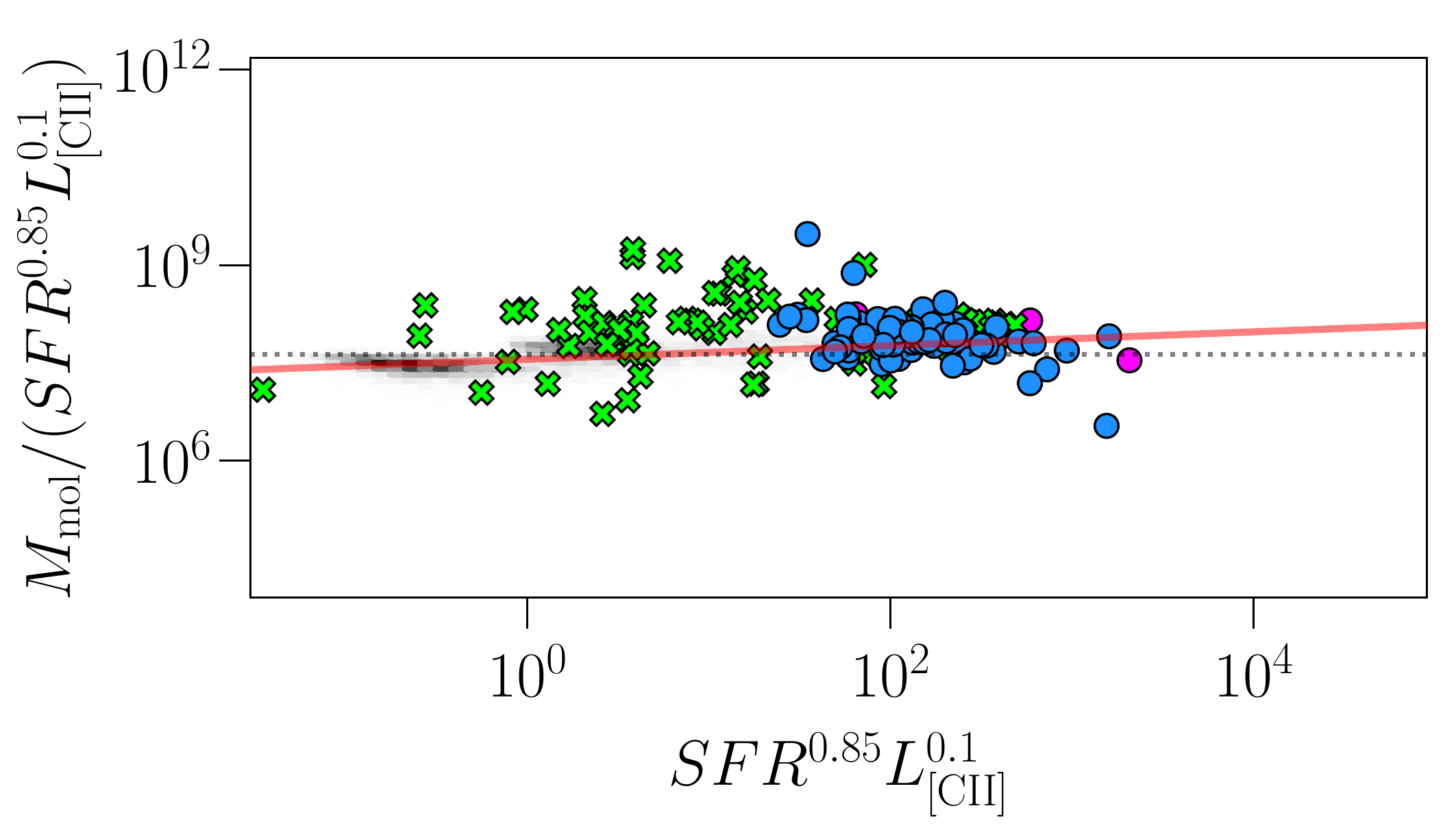}
     \caption{Equation \ref{eq:pca} (top) and \ref{eq:pca_b} (bottom) for our simulated galaxies, ALPINE galaxies, and Z18 galaxies. }
     \label{figure:pca}
 \end{figure}
The unobscured star formation rate can be measured
in a number of ways, from the luminosity of the rest-frame continuum at
$1500\,{\rm \AA}$, for example, or from the Ly$\alpha$ emission.
Obscured star formation rates are typically inferred from (sub-)millimeter or
far-IR continuum measurements.  We consider here the total star formation rates
of our simulated galaxies and do not distinguish between obscured or
unobscured star formation rates.
We will also examine whether including the stellar masses might lead to more accurate gas mass estimates.
To this end, we perform a principal component analysis (PCA) on the parameter space spanned by $\log L_{\rm [CII]}$, $\log M_{\rm mol}$,
$\log {\rm SFR}$, and $\log M_{\rm \star}$. This gives us the following principal components:

\tiny
\begin{eqnarray}
{\rm PC1} &=& 0.47 \log L_{\rm [CII]} + 0.51 \log {\rm SFR} + 0.51 \log M_{\rm mol} + 0.51 \log M_{\rm \star}\\
{\rm PC2} &=& -0.87 \log L_{\rm [CII]} + 0.40 \log {\rm SFR} + 0.26 \log M_{\rm mol} + 0.14 \log M_{\rm \star}\\
{\rm PC3} &=& -0.14 \log L_{\rm [CII]} -0.41 \log {\rm SFR} -0.31 \log M_{\rm mol} + 0.85 \log M_{\rm \star}\\
{\rm PC4} &=& 0.08 \log L_{\rm [CII]} + 0.64 \log {\rm SFR} -0.76 \log M_{\rm mol} + 0.05 \log M_{\rm \star}.
\end{eqnarray}
\normalsize

Of the four principal components, PC1 is responsible for 91\% of the variance, and PC2 and PC3 are responsible for 7 and 2\%, respectively.
Given its negligible contribution to the variance, we can set PC4 to zero, which in turn allows us to derive the following expression for the molecular gas:
\begin{equation}
    \log M_{\rm mol} = 0.10 \log L_{\rm [CII]} + 0.85  \log SFR + 0.07 \log M_{\rm \star}.
\label{eq:pca}
\end{equation}
The dominant principal component in this prescription is the star formation rate, as expected given that star formation is fueled by molecular gas.
The [C{\sc ii}] luminosity and stellar mass also account for some of the correlation with molecular gas mass, 
each with about 10\% of the contribution. We thus interpret the $\log L_{\rm \cii}- \log M_{\rm mol}$ relation at high-$z$ as a second-order result of the Kennicutt-Schmidt Law, where \cii luminosity is a coincidental tracer of the molecular gas mass as a result of the \cii-SFR relation.

In Fig.\ \ref{figure:pca} we have plotted
eq.\ \ref{eq:pca} for our simulations and comparison data,  with (top) and without the $M_{\rm \star}$-term (bottom). The residual scatter in the two cases is virtually identical, i.e., $0.18$ and $0.19\,{\rm dex}$, respectively, and we can therefore drop $M_{\rm \star}$, which is the most difficult observable from eq.\,\ref{eq:pca}.   Thus, a useful relation to obtain the molecular gas mass to within a fractional scatter of $0.19 \ln(10)\simeq 44\%$ is: 
\begin{equation}
    \log M_{\rm mol} = 0.10 \log L_{\rm [CII]} + 0.85  \log \rm{SFR}.
    \label{eq:pca_b}
\end{equation}
The conclusion from our PCA analysis is that the $\log L_{\rm [CII]} - \log M_{\rm mol}$ relation comes about from
a combination of the $\log L_{\rm [CII]} - \log {\rm SFR}$ relationship and the Kennicutt-Schmidt relation, which relates ${\rm SFR}$ to $M_{\rm mol}$.

\citet{Sommovigo2021} came to a similar conclusion in that they provided a plausible physical 
explanation for the invariant $\alpha_{\rm [CII]}$ by combing the $\Sigma_{\rm [CII]} - \Sigma_{\rm SFR}$ relation \citep{delooze2014}, and the $\Sigma_{\rm SFR} - \Sigma_{\rm mol}$ \citep{Kennicutt1998, kennicutt2012}. 
They found that $\alpha_{\rm [CII]} = 30.3\tau_{\rm depl} \Sigma_{\rm SFR}^{-0.075}$, where $\tau_{\rm depl} = M_{\rm mol}/{\rm SFR}$ is the molecular gas depletion time-scale in Gyr, and $\Sigma_{\rm SFR}$ is the star-formation rate surface density in ${\rm \Msolar\,yr^{-1}\,kpc^{-2}}$.
Given the weak dependence on 
$\Sigma_{\rm SFR}$ and observational indications that for main sequence galaxies, $\tau_{\rm depl}$ is approximately constant ($\sim 0.4-0.7\,{\rm Gyr}$) across the redshift range $z\sim 0 - 2$
they find a nearly constant value of $\alpha_{\rm [CII]} = (12-21)\Sigma_{\rm SFR}^{-0.075}$, which is in good agreement with the Z18 values ($\simeq 30\,{\rm M_{\rm \odot}/L_{\rm \odot}}$) as well as the $\alpha_{\rm [CII]}$ values for our $z=6$ simulations ($\simeq 18\,{\rm M_{\rm \odot}/L_{\rm \odot}}$).

\begin{figure}
 \centering
 \includegraphics[width=\columnwidth]{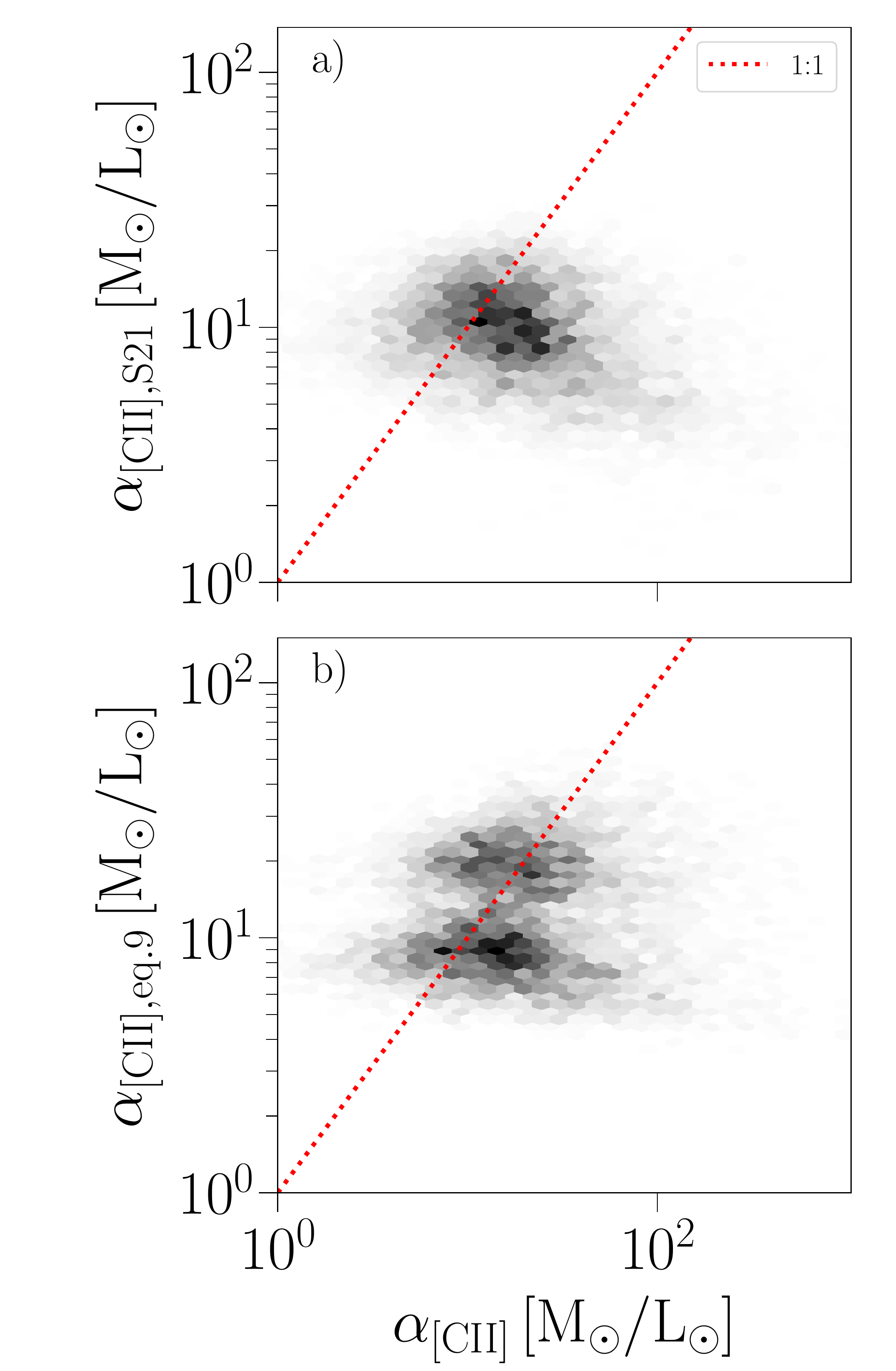}
     \caption{$\alpha_{\rm [CII]}$-values calculated for our simulations using the relation from \citet{Sommovigo2021} (panel {\bf a)}) and eq.~\ref{eq:our-eq} (panel {\bf b)}) vs the true $\alpha_{\rm [CII]} = M_{\rm mol}/L_{\rm [CII]}$ values for our simulations. The red dotted lines denotes the 1:1 relationship. }
     \label{fig:alpha_S21_sims}
 \end{figure}

Following \citet{Sommovigo2021}, we derive an integrated version of the above relation. To this end, we adopt ${\rm SFR} = 10^{-8.52}L_{\rm \cii}^{1.18}$, which \cite{delooze2014} found applied to $z > 0.5$ starburst galaxies. For the integrated Kennicutt-Schmidt relation we adopt 
$M_{\rm mol} = 10^{8.00}{\rm SFR^{0.83}}$, which \citet{Sargent2014} found applied to starbursts. Combining the two, we derive:
\begin{equation}
    \alpha_{\rm \cii} = 60.3 \ \tau_{\rm depl} \ {\rm SFR}^{0.16},
\label{eq:our-eq}
\end{equation}
where $\tau_{\rm depl}$ is in Gyr. This relation is similar to that of \citet{Sommovigo2021}, both in terms of its normalization and the weak ${\rm SFR}$-dependence, albeit with a positive exponent rather than a negative one. 
In Fig.~\ref{fig:alpha_S21_sims} we plot $\alpha_{\rm \cii}$ for our simulations derived from the \citet{Sommovigo2021} relation (top) and eq.~\ref{eq:our-eq} (bottom) vs their $\alpha_{\rm \cii} = M_{\rm mol}/L_{\rm \cii}$ values.  
It is seen that neither the $\alpha_{\rm \cii}$ estimator provided by \citet{Sommovigo2021} nor its volumetric version derived here correlate well with the  $\alpha_{\rm \cii} = M_{\rm mol}/L_{\rm \cii}$ values of the simulations, and significant scatter around the 1:1 relation, is seen along the horizontal direction. However, both estimators agree in the mean with the simulation values.
The apparent bimodality seen in Fig.\,\ref{fig:alpha_S21_sims}b is due to the fact that our simulations, which come from three simulation boxes (see \S\ref{section:simulations}) span a wide range in both SFR and $\tau_{\rm depl}$  (Fig.\,\ref{fig:main_sequence} top and bottom panels). The bottom cloud of points in Fig.\,\ref{fig:alpha_S21_sims}b is predominantly from the low-SFR and low-$\tau_{\rm depl}$ galaxies extracted from the $25\,{\rm cMpc}\,h^{-1}$ box, while the top cloud consists of galaxies from the $50$ and $100\,{\rm cMpc}\,h^{-1}$ box. This bimodality is not seen in  Fig.\,\ref{fig:alpha_S21_sims}a since i) the dependency on $\Sigma_{\rm SFR}$ in the relation by \citet{Sommovigo2021}is significantly weaker and ii) the simulations span a relatively narrow range in $\Sigma_{\rm SFR}$.

Extreme starburst galaxies have molecular gas depletion time-scales of $\sim 100\,{\rm Myr}$ \citep[e.g.,][]{Aravena2016}, i.e., of the order $4-7\times$ shorter than main-sequence galaxies, which would imply correspondingly lower $\alpha_{\rm \cii}$-values for starbursts (see eq.~\ref{eq:our-eq}). The $\sim 5-10\times$ higher star-formation rates of starburst galaxies is not sufficient to fully offset the effect of the shorter gas-depletion time-scales on $\alpha_{\rm \cii}$. Assuming that the gas depletion time-scales and star-formation rates for the main-sequence galaxies (MS) and starburst galaxies (SB) roughly scale as $\tau_{\rm SB} \sim (0.4-0.7)\times \tau_{\rm MS}$ and ${\rm SFR_{\rm SB} \sim (5-10)\times SFR_{\rm MS}}$, we find from eq.~\ref{eq:our-eq} that $\alpha_{\rm \cii, SB} \sim (0.2-0.4)\times \alpha_{\rm \cii, MS}$. Inserting our median value of $\alpha_{\rm \cii} = 18\,{\rm M_{\rm \odot}/L_{\rm \odot}}$ from our simulations, then yields an expected conversion factor for starbursts of $\alpha_{\rm \cii, SB} \sim (4-7)\,{\rm M_{\rm \odot}/L_{\rm \odot}}$. This is consistent with the low values found in high-$z$ dusty starburst galaxies by \citet{rizzo2021} (see \S\ref{subsection:alpha-cii}).

\section{Conclusion}

Using S\'IGAME postprocessing of galaxies from \simba cosmological simulations, 
we have simulated and analysed the \cii emission from 11,129 $z = 6$ galaxies with the goal of examining the viability of \cii as a tracer of the ISM gas mass, and in particular the molecular gas mass, in normal star-forming galaxies at this epoch.
The \simba simulated galaxies span the stellar mass range $1.5\times 10^7-5\times 10^{10}\,{\rm \Msolar}$
and define a star-forming main sequence at $z = 6$ that at the high-mass-end agrees with observations, and at the low-mass-end agrees with extrapolations of the observed main-sequence \citet{Schreiber2015}. 
The simulated galaxies also follow the empirical Kennicutt-Schmidt law and the mass-metallicity relation.
Our main findings are presented below:

\begin{enumerate}
    \item 
    For the most massive ($M_{\rm \star} \gs 10^9\,{\rm \Msolar}$) galaxies in our simulation, the \cii emission is almost entirely coming from the molecular gas phase. For galaxies with lower stellar masses ($M_{\rm star} = 10^7-10^9\,{\rm \Msolar}$) there is a larger spread in the fraction ($\sim 50-100\%$) of the total \cii emission coming from the molecular gas, but the general trend is that the average fraction decreases towards lower masses. The atomic gas phase contributes a negligible amount ($\ls 5\%$) to the total \cii emission. As a result, the contribution from the ionized ISM to the \cii emission is the mirror opposite to that of the molecular gas.  
    
    \item We find and parametrize a log-linear correlation between the \cii emission and molecular gas masses of our galaxies: ${\log L_{\rm \cii} = (0.78\pm 0.01) \log M_{\rm mol} + (0.44\pm 0.04)}$. The slope is shallower than the near-unity slope derived by Z18, and the scatter in the correlation is $0.45\,{\rm dex}$, which is larger than the $0.2\,{\rm dex}$ scatter in the Z18 relation. The shallower slope is due the simulated galaxies at the low-mass end ($\ls 10^8\,{\rm \Msolar}$) generally having higher \cii luminosities compared to the extrapolation of the linear Z18 relation to these low masses. However, the observations that are available at the low-mass end \citep[with the exception of][]{madden2020} are consistent with the simulation predictions, albeit with a larger scatter. Thus, the difference in slope might be attributable to a large amount of dwarf galaxies in our sample which lack comparable observational data at high redshifts. At the high-mass end, our simulations are in good agreement with the Z18-relation and observational data from Z18 and ALPINE. 

    \item We derive a \cii-to-$M_{\rm mol}$ conversion factor, $\alpha_{\rm \cii}$, based on our simulations, finding a median value of $\alpha_{\rm \cii} = 18$ \msun{}/\lsun{} with a median absolute deviation of 10 \msun{}/\lsun{}. This is lower than the average conversion factor derived by Z18, although there is a significant overlap in the distributions between our simulated and the observed $\alpha_{\rm \cii}$-values. We attribute the lower average $\alpha_{\rm \cii}$-values in our simulations to their much lower gas-phase metallicities, which allow for a higher CO photo-dissociation rate and for neutral carbon atoms in the ISM to be more readily ionized. 
    
    \item Using principal component analysis, we determine a relation 
    between molecular gas mass, star formation rate, and \cii emission, ${\log M_{\rm mol} = 0.10 \log L_{\rm [CII]} + 0.85 \log \rm{SFR}}$, which has a scatter of only $0.18\,{\rm dex}$. Our analysis suggests that the
    tight relationship between the star-formation rate and the molecular gas mass is ultimately responsible for \cii as a tracer of molecular gas mass. 
    
\end{enumerate}

\subsection{Acknowledgements}

We thank the referee for valuable insight and constructive feedback which helped to significantly improve the results of this paper. We thank Robert Thompson for developing \caesar, and the \yt team
for development and support of \yt. 
This research was made possible by the National Science Foundation (NSF)-funded DAWN-IRES program 
and would not be possible without support from the Cosmic Dawn Center (DAWN) in Copenhagen, Denmark. 
The Cosmic Dawn Center (DAWN) is funded by the Danish National Research Foundation under grant No. 140.
DV is funded by an Open Study/Research Award from the Fulbright U.S. Student Program in Denmark.
KPO is funded by NASA under award No 80NSSC19K1651. GEM acknowledges  the Villum Fonden research grant 37440, "The Hidden Cosmos” and the Cosmic Dawn Center of Excellence funded by the Danish National Research Foundation under then grant No. 140. 
DN acknowledges support from the NSF via grant AST 1909153. 
RD acknowledges support from the Wolfson Research Merit Award program of the U.K. Royal Society.
KEH acknowledges support by a Postdoctoral Fellowship Grant (217690--051) from The Icelandic Research Fund and from the Carlsberg Foundation Reintegration Fellowship Grant CF21-0103.
\simba was run on the DiRAC@Durham facility managed by the Institute for Computational Cosmology on behalf of the STFC DiRAC HPC Facility. The equipment was funded by BEIS capital funding via STFC capital grants ST/P002293/1, ST/R002371/1 and ST/S002502/1, Durham University and STFC operations grant ST/R000832/1. DiRAC is part of the National e-Infrastructure.

\bibliography{bibs}

\end{document}